\documentclass[sts]{imsart}

\RequirePackage[OT1]{fontenc}
\usepackage{amsthm,amsmath,amsfonts,natbib,graphicx}
\usepackage{epstopdf}

\RequirePackage[colorlinks,citecolor=blue,urlcolor=blue]{hyperref}

\startlocaldefs
\numberwithin{equation}{section}
\theoremstyle{plain}

\newtheorem{lemma}{Lemma}[section]
\endlocaldefs

\makeatletter
\newcommand*{\centernot}{%
  \mathpalette\@centernot
}
\def\@centernot#1#2{%
  \mathrel{%
    \rlap{%
      \settowidth\dimen@{$\m@th#1{#2}$}%
      \kern.5\dimen@
      \settowidth\dimen@{$\m@th#1=$}%
      \kern-.5\dimen@
      $\m@th#1\not$%
    }%
    {#2}%
  }%
}
\makeatother

\newcommand{\independent}{\perp\mkern-9.5mu\perp}
\newcommand{\notindependent}{\centernot{\independent}}

\begin{document}

\begin{frontmatter}
\title{The GENIUS Approach to Robust Mendelian Randomization Inference}
\runtitle{Robust Mendelian Randomization Inference}
%\thankstext{T1}{Footnote to the title with the `thankstext' command.}

\begin{aug}
\author{\fnms{Eric} \snm{Tchetgen Tchetgen}\ead[label=e1]{ett@wharton.upenn.edu}},
\author{\fnms{BaoLuo} \snm{Sun}\ead[label=e2]{stasb@nus.edu.sg}}
\and
\author{\fnms{Stefan} \snm{Walter}
\ead[label=e3]{swalter@psg.ucsf.edu}}

%\thankstext{t1}{Some comment}
%\thankstext{t2}{First supporter of the project}
\runauthor{Tchetgen Tchetgen et al.}

%\affiliation{Some University and Another University}

\address{Eric Tchetgen Tchetgen is Professor, Department of Statistics, The Wharton School of the University of Pennsylvania, U.S.A. \printead{e1}. BaoLuo Sun is Assistant Professor, Department of Statistics and Applied Probability, National University of Singapore, Singapore \printead{e2}. Stefan Walter is affiliated with the Department of Epidemiology and Biostatistics, University of California San Francisco, U.S.A. \printead{e3}.
}

\end{aug}

\begin{abstract}
Mendelian randomization (MR) is a popular instrumental variable (IV)
approach, in which one or several genetic markers serve as IVs that can
sometimes be leveraged to recover valid inferences about a given
exposure-outcome causal association subject to unmeasured confounding. A key
IV identification condition known as the exclusion restriction states that the
IV cannot have a direct effect on the outcome which is not mediated by the
exposure in view. In MR studies, such an assumption requires an unrealistic
level of prior knowledge about the mechanism by which genetic markers causally
affect the outcome. As a result, possible violation of the exclusion
restriction can seldom be ruled out in practice. To address this concern, we
introduce a new class of IV estimators which are robust to violation of the
exclusion restriction under data generating mechanisms commonly assumed in MR
literature. The proposed approach named "MR G-Estimation under No Interaction
with Unmeasured Selection" (MR GENIUS) improves on Robins' G-estimation by
making it robust to both additive unmeasured confounding and violation of the
exclusion restriction assumption. In certain key settings, MR GENIUS reduces to the estimator of \cite{doi:10.1080/07350015.2012.643126} which is widely used in econometrics but appears largely unappreciated in MR literature. More generally, MR GENIUS generalizes Lewbel's estimator to several key practical MR settings, including multiplicative causal models for binary outcome, multiplicative and odds ratio exposure models, case control study design and censored survival outcomes.
\end{abstract}

\begin{keyword}
\kwd{additive model}
\kwd{confounding}
\kwd{exclusion restriction}
\kwd{G-estimation}
\kwd{instrumental variable}
\kwd{robustness}
\end{keyword}

\end{frontmatter}

\section{Introduction}
\label{s:intro}

Mendelian randomization (MR) is an instrumental variable approach with growing
popularity in epidemiology studies. In MR, one aims to establish a causal
association between a given exposure and an outcome of interest in the
presence of possible unmeasured confounding, by leveraging one or more genetic
markers defining the IV \citep{davey2003mendelian,10.1093/ije/dyh132, doi:10.1002/sim.3034}. In order to be a valid IV, a genetic marker must satisfy the following
key conditions:

\begin{description}
\item[(a)] It must be associated with the exposure.

\item[(b)] It must be independent of any unmeasured confounder of the
exposure-outcome relationship.

\item[(c)] There must be no direct effect of the genetic marker on the outcome
which is not fully mediated by the exposure in view.
\end{description}

Assumption (c) also known as the exclusion restriction is rarely credible in
the context of MR as it requires complete understanding of the biological
mechanism by which each marker influences the outcome. Such a priori knowledge
may be unrealistic in practice due to the possible existence of unknown
pleiotropic effects of the markers \citep{LITTLE2003930, davey2003mendelian,10.1093/ije/dyh132, doi:10.1002/sim.3034}. Violation of assumption (b) can also
occur due to linkage disequilibrium or population stratification \citep{ doi:10.1002/sim.3034}. Possible violation or near violation of assumption (a) known as the
weak instrumental variable problem also poses an important challenge in MR\ as
individual genetic effects on phenotypes can be fairly weak.

There has been tremendous interest in the development of statistical methods
to detect and account for violation of IV assumptions (a)--(c), primarily in
multiple-IV settings under standard linear outcome and exposure models. The
literature addressing violation of assumption (a) is arguably the most
developed and extends to possible nonlinear models under a generalized methods
of moments framework; notable papers of this rich literature include \cite{10.2307/2171753}, \cite{doi:10.1111/1468-0262.00151}, \cite{10.2307/1392421}, \cite{10.2307/3598886} and \cite{newey2009generalized}. Methodology to address violations of (b) or (c)\ is far less
developed, and constitutes the central focus of this paper. Three strands of
work stand out in recent literature concerning violation of either of these
assumptions. In the first strand, \cite{doi:10.1080/01621459.2014.994705} developed a penalized
regression approach that can recover valid inferences about the causal effect
of interest provided fewer than fifty percent of genetic markers are invalid
IVs (known as majority rule); \cite{doi:10.1080/01621459.2018.1498346} improved on the
penalized approach, including a proposal for standard error estimation lacking
in \cite{doi:10.1080/01621459.2014.994705}. In an alternative approach, \cite{HAN2008285} established that
the median of multiple\ estimators of the effect of exposure obtained using
one instrument at the time is a consistent estimator also assuming majority
rule and that IVs cannot have direct effects on the outcome unless the IVs are
uncorrelated. \cite{doi:10.1002/gepi.21965} explore closely related weighted median
methodology. In a second strand of work, \cite{doi:10.1111/rssb.12275} proposed two stage
hard thresholding (TSHT) with voting, which is consistent for the causal
effect under linear outcome and exposure models, and a plurality rule which
can be considerably weaker than the majority rule. The plurality rule is
defined in terms of regression parameters encoding the association of each
invalid IV with the outcome and that encoding the association of the
corresponding IV\ with the exposure. The condition effectively requires that
the number of valid IVs is greater than the largest number of invalid IVs with
equal ratio of the regression coefficients given above. Furthermore, they provide a
simple construction for 95\% confidence intervals to obtain inferences about
the exposure effect which are guaranteed to have correct coverage under the
plurality rule. Importantly, in these first two strands of work, a candidate
IV may be invalid either because it violates the exclusion restriction, or
because it shares an unmeasured common cause with the outcome, i.e. either (b)
or (c) fails. Both the penalized approach and the median estimator may be
inconsistent if majority rule fails, while TSHT may be inconsistent if
plurality rule fails. For instance, it is clear that neither approach can
recover valid inferences if all IVs violate either assumption (b) or (c). In a
third strand of work, \cite{doi:10.1080/07350015.2014.978175} considered the possibility of
identifying the exposure causal effect when all IVs violate the exclusion
restriction (c), provided the effects of the IVs on the exposure are
asymptotically orthogonal to their direct effects on the outcome as the number
of IVs tends to infinity. A closely related meta-analytic version of their
approach known as MR-Egger has recently emerged in the epidemiology literature
\citep{bowden2015mendelian}; they referred to the orthogonality condition as the
instrument strength independent of direct effect (InSIDE) assumption. As
pointed out by \cite{doi:10.1080/01621459.2014.994705}, the orthogonality condition on which these
approaches rely may be hard to justify in MR settings as it potentially
restricts unknown pleiotropic effects of genetic markers often with little to
no biological basis. A notable feature of aforementioned methods is that they
are primarely tailored to a multiple-IV setting, in fact methods such as
MR-Egger are consistent only under an asymptotic theory in which the number of
IVs goes to infinity, together with sample size. It is also important to note
that because confidence intervals for the causal effect of the exposure
obtained by \cite{doi:10.1080/01621459.2018.1498346} and \cite{doi:10.1111/rssb.12275} rely on a consistent
model selection procedure, such confidence intervals fail to be uniformly
valid over the entire model space \citep{doi:10.1111/rssb.12275, LEEB2008201}.

Because in practice, it is not possible to ensure that either majority rule or
plurality rule holds, it is important to develop causal inference and estimation methods that are robust to possible violation of IV assumptions under alternative conditions. \cite{doi:10.1080/07350015.2012.643126, LEWBEL201810} proposed novel identification and estimation strategies with mismeasured and endogenous regressor models by leveraging a heteroscedastic covariance restriction, which has since been widely applied in econometrics and social sciences. In Section \ref{s:not}, we introduce notation used throughout and provide a review of the invalid IV model considered by \cite{doi:10.1080/07350015.2012.643126}. We extend Lewbel's identification result in section \ref{s:est}. The proposed framework which we call "MR G-Estimation under No
Interaction with Unmeasured Selection" (MR GENIUS) can also be viewed as a version
of Robins` G-estimation \citep{doi:10.1080/03610929408831393} that is robust to both additive unmeasured confounding and violation of IV assumptions, and which unlike the aforementioned methods equally applies whether one has observed a single or many candidate IVs. An important feature of multiple IV MR
GENIUS is that the correlation structure for the IVs can essentially remain
unrestricted without necessarily affecting identification, this is in contrast
with \cite{bowden2015mendelian} who require uncorrelated IVs and \cite{doi:10.1080/01621459.2014.994705}
who likewise require IV correlation structure to be somewhat restricted
\citep{doi:10.1080/01621459.2018.1498346}.  Section \ref{s:gen} presents several key generalizations including MR
GENIUS under multiplicative or odds ratio exposure models, as well as for
right censored time-to-event endpoint under a structural additive hazards
model, which extends the recent semiparametric IV estimator of
\cite{doi:10.1111/biom.12699} against possible violation of the exclusion
restriction assumption. In section \ref{s:sim}, we evaluate the proposed methods and
compare them to a number of previous MR methods in extensive simulation
studies. In section \ref{s:app} we illustrate the methods in an MR analysis of the
effect of diabetes on memory in the Health and Retirement Study. Section \ref{s:dis}
offers some concluding remarks. 

\section{Notation and definitions}
\label{s:not}
Suppose that one has observed $n$ i.i.d. realizations of a vector $(A,G,Y)$
where $A$ is an exposure, $G$ the candidate IV and $Y$ is the outcome. Let $U$
denote an unmeasured confounder (possibly multivariate) of the effect of $A$
on $Y.$ $G$ is said to be a valid instrumental variable provided it fulfills
the following three conditions \citep{didelez2010}:

\begin{description}
\item[Assumption 1. ] IV\ relevance: $G\notindependent A|U;$

\item[Assumption 2. ] IV independence: $G\independent U;$

\item[Assumption 3. ] Exclusion restriction: $G\independent Y|A,U.$
\end{description}

\begin{figure}
\begin{center}
\centerline{\includegraphics[width=0.9\textwidth]{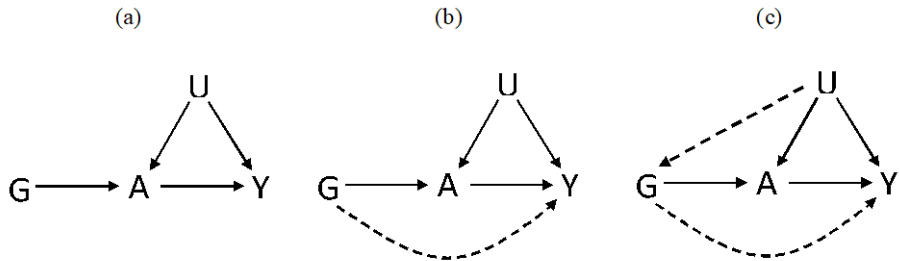}}
\end{center}
\caption{Directed acyclic graph
depicting (a) a valid instrument $G$ which satisfies assumptions 1-3, (b) the situation in which exclusion restriction (assumption 3) does not
necessarily hold (the dashed line indicates possible direct effect of $G$ on
outcome $Y$), and (c) the situation in which IV independence (assumption 2) and exclusion
restriction (assumption 3) do not necessarily hold (the dashed lines indicate
possible direct effects of $U$ on $G$, and of $G$ on $Y$).\label{fig1}}
\end{figure}

\noindent The first condition ensures that the IV\ is a correlate of the
exposure even after conditioning on $U.$ The second condition states that the
IV\ is independent of all unmeasured confounders of the exposure-outcome
association, while the third condition formalizes the assumption of no direct
effect of $G$ on $Y$ not mediated by $A$ (assuming Assumption 2 holds$)$.
\ The causal diagram in Figure \ref{fig1}a encodes these three assumptions and therefore
provides a graphical representation of the IV\ model. It is well known that
while a valid IV satisfying assumptions 1--3, i.e. the causal diagram in Figure
\ref{fig1}a, suffices to obtain a valid statistical test of the sharp null hypothesis of
no individual causal effect, the population average causal effect is itself
not point identified with a valid IV without an additional assumption. In case of a valid binary IV and binary
exposure, \cite{doi:10.1111/rssb.12262} recently established that the
average causal effect $\beta_{a}$ of $A$ on $Y$ is nonparametrically identified by the
so-called Wald estimand $$\beta_{a}=\delta\equiv\frac{E(Y|G=1)-E(Y|G=0)}{E(A|G=1)-E(A|G=0)},$$ if either of the conditions
\begin{eqnarray}
\label{eq:wangeric}
E\left(  Y|A,G,U\right)     =\beta_{a}A +\beta_{u}\left(  U\right)\\
E\left(  A|G,U\right)   =\alpha_{g}G  +\alpha_{u}(U),
\end{eqnarray}
holds, where the
unknown functions $\beta_{u}\left(  \cdot\right)  $ and $\alpha_{u}(\cdot)$ are restricted only
by natural features of the model, e.g. such that the outcome and
exposure means are bounded between zero and one in the binary case. Suppose that, as encoded in the diagram given in Figure \ref{fig1}b, assumption 3 does not necessarily hold. \cite{doi:10.1080/07350015.2012.643126} considered identification and estimation of $\beta_a$ under the invalid IV model
\begin{eqnarray}
\label{eq:lewbel}
E\left(  Y|A,G,U\right)     &=&\beta_{a}A + \beta_{g}(G)+ \beta_{u}\left(  U\right)\\
 \quad E\left(  A|G,U\right)   &=&\alpha_{g}(G)  +\alpha_{u}(U) \label{exp},
\end{eqnarray}
where $\beta_{g}(G)$ encodes the direct effect of $G$ on $Y$, with $ \beta_{g}(0)=\alpha_g(0)=0$. Note that the models for $E(Y|A,G,U)$ and $E(A|G,U)$ considered by \cite{bowden2015mendelian} satisfy (\ref{eq:lewbel}) with $\beta_{g}\left(
\cdot\right)  $ and $\alpha_{g}\left(  \cdot\right)  $ linear functions, while
\cite{doi:10.1080/01621459.2014.994705} specified models implied by these two restrictions. With binary exposure $A$, we consider identification of the average causal effect for the following generalization of the invalid IV model considered in (\ref{eq:lewbel}).

\begin{description}
\item[Assumption 4:]
\item[(4a)] For unknown functions $\beta_{g}\left(  \cdot\right)$  and $\alpha_{g}(\cdot)$ of $(U,G)$ where $\beta_{g}\left(  U,0\right)=\alpha_{g}(U,0)=0$, and unknown function $\beta_a(\cdot)$ of $U$,
\begin{eqnarray}
E\left(  Y|A,G,U\right)     &=&\beta_{a}(U)A + \beta_{g}(U,G)+ \beta_{u}\left(  U\right)\\
 E\left(  A|G,U\right)   &=&\alpha_{g}(U,G)  +\alpha_{u}(U).
\label{4.a1}%
\end{eqnarray}

\item[(4b)] The orthogonality conditions
\begin{eqnarray}
\label{orth}
&&\mbox{cov}\{\beta_{g}\left(  U,G\right),\alpha_g(U,G)|G\}=\mbox{cov}\{\beta_{g}\left(  U,G\right),\alpha_u(U)|G\}=0;\notag \\
&&\mbox{cov}\{\beta_a(U), \alpha_g(U,G)|G\}=\mbox{cov}\{\beta_a(U), \alpha_u(U)|G\}=0;\\
&&\mbox{cov}\{\alpha_g(U,G),\beta_{u}\left(  U\right)|G\}=0, \notag
\end{eqnarray}
hold with probability 1.
\end{description}

Under assumption 4a, the average causal effect of binary $A$ on $Y$ is given by $E\{\beta_{a}(U)\}$. The model $\beta_{g}(U,G)$ encodes the direct effect of $G$ on $Y$, with potential effect modification by unmeasured confounders $U$. Assumption 4b does not imply orthogonality of $\beta_u(U)$ and $\alpha_u(U)$ and therefore the degree of unmeasured confounding is not restricted by these orthogonality conditions. In contrast the degree of common effect modifiers in the outcome and exposure models is effectively restricted by (\ref{orth}). As a special case, the conditions in assumption 4b are satisfied if $\beta_g(U,G)=\beta_g(G)$, $\alpha_g(U,G)=\alpha_g(G)$ and $\beta_a(U)=\beta_a$ with probability 1, which is the scenario considered in \cite{doi:10.1080/07350015.2012.643126}.  Assumption 4b may hold even if assumptions made by \cite{doi:10.1080/07350015.2012.643126} do not, as illustrated by the following example. Suppose that $\alpha_u(U)-E[\alpha_u(U)]=\sum_{k=1}^K \gamma_k \times \phi_k(U)$  for a vector of zero-mean functions $\mathbf{\Phi}=\{\phi_k:k=1,..K\}$, and likewise $\beta_u(U)-E[\beta_u(U)] =\sum_{j=1}^J \theta_j\times \tau_j(U)$ for $\mathbf{T}=\{\tau_j:j=1,..J\}$. Denote the linear vector space spanned by the vector $\mathbf{\Omega}$ of functions in $U$ to be $S(\mathbf{\Omega})$, and $\Pi$ to be the least squares projection operator.  Let $$\alpha_g(U,G)=\alpha_{g0}(G)+\sum_{k=1}^K  \alpha_{gk}(G)\{\phi_k(U)-\Pi[\phi_k(U)|S(\mathbf{T})]\},$$ $\alpha_{gk}(0)=0$ for $k=0,1,...,K$, 
\begin{eqnarray*}
&&\beta_g(U,G)=\beta_{g0}(G)\\
&+&\beta_{g1}(G)\{\omega(U)-\Pi[\omega(U)|S(\mathbf{\Phi})+ S(\phi_k(U)-\Pi[\phi_k(U)|S(\mathbf{T})],k=1,2,...,K)]\},
\end{eqnarray*}
for arbitrary function $\omega(U)$ with $\beta_{g0}(0)=\beta_{g1}(0)=0$, and $$\beta_a(U)=\zeta(U)-\Pi[\zeta(U)|S(\mathbf{\Phi})+ S(\phi_k(U)-\Pi[\phi_k(U)|S(\mathbf{T})], k=1,2,...,K)]$$ for arbitrary function $\zeta(U)$. Note that $\mathbf{\Phi}$ and $\mathbf{T}$ can be of different dimension. Then the orthogonality conditions of Assumption 4b are satisfied.

\section{Identification and estimation under violation of exclusion restriction}
\label{s:est}

 We consider identification of $E\{\beta_{a}(U)\}$ within the large class of data generating mechanisms that satisfy assumptions 1, 2 and 4, which is given in the following lemma.

\begin{lemma}
\label{l1}
\qquad Suppose that Assumptions 1, 2 and 4 hold, then $E\{\beta_{a}(U)\}=\mu$, where
\begin{eqnarray*}
\mu &  \equiv&\frac{E\left[  \left\{  G-E(G)\right\}  \left\{  A-E(A|G)\right\}
Y\right]  }{E\left[  \left\{  G-E(G)\right\}  \left\{  A-E(A|G)\right\}
A\right]  }
\end{eqnarray*}
provided that
\begin{equation}
\phi\equiv\mbox{cov} \{G,\mbox{var}(A|G)\}\neq 0.  \label{strong IV}%
\end{equation}
\end{lemma}
The proof for Lemma \ref{l1} is given in appendix A1 of the supplementary materials. In particular, for binary $G$, $\phi=var(G)\left\{  var(A|G=1)-var(A|G=0)\right\}$ so that (\ref{strong IV}) is satisfied if and only if
$\mbox{var}\left(  A|G=1\right)  -\mbox{var}\left(  A|G=0\right)  \neq0$. Lemma 1 provides an explicit
identifying expression for the average causal effect of $A$ on $Y$
in the presence of additive confounding, which leverages a candidate IV $G$
that may or may not satisfy the exclusion restriction. In order for $\mu$ to
be well defined, we require a slight strengthening of the IV relevance
assumption 1, i.e. that $var(A|G)$ must depend on $G$. It is key to note that
this assumption is empirically testable, and will typically hold for binary
$A$, other than at certain exceptional laws. To illustrate, let $\pi
(g)=\Pr(A=1|G=g)$ and suppose that assumptions 1, 2 and 4 hold, however
$\pi\left(  1\right)  =1-\pi\left(  0\right)  ,$ in which case $\left(
\ref{strong IV}\right)  $ fails because $var(A|G=g)=\pi(g)\left(
1-\pi(g)\right)  =\pi(1)\left(  1-\pi(1)\right)  =\pi(0)\left(  1-\pi
(0)\right)  $ does not depend on $g$ and therefore the identifying expression
given in the Lemma does not apply despite the candidate IV satisfying IV
relevance assumption 1, i.e. $\pi\left(  1\right)  \not =\pi\left(  0\right)
$. Below, we extend Lemma \ref{l1} to allow for possible violation of both
assumptions 2 and 3.

The lemma motivates the following MR\ estimator, which is guaranteed to be
consistent under assumptions 1, 2, 4 and equation $\left(  \ref{strong IV}%
\right)  $ irrespective of whether or not assumption 3 also holds:%
\begin{equation}
\widehat{\beta}_{a}=\frac{\mathbb{P}_{n}\left[  \left\{  G-\mathbb{P}%
_{n}(G)\right\}  \left\{  A-\widehat{E}(A|G)\right\}  Y\right]  }%
{\mathbb{P}_{n}\left[  \left\{  G-\mathbb{P}_{n}(G)\right\}  \left\{
A-\widehat{E}(A|G)\right\}  A\right]  }, \label{Estimator}%
\end{equation}
where $\mathbb{P}_{n}=n^{-1}\sum_{i=1}^{n}[\cdot]_{i}$ and $\widehat{E}%
(A|G=g)=$ $\mathbb{P}_{n}\left[  A_{i}1\left(  G_{i}=g\right)  \right]
/\mathbb{P}_{n}\left[  1\left(  G_{i}=g\right)  \right]  .$ This estimator is
the simplest instance of MR GENIUS estimation. The asymptotic distribution of
the estimator is described in appendix B1. We note that  (\ref{Estimator}) is equivalent to Lewbel's estimator which can be implemented as follows \citep{LEWBEL201810}: 
\begin{itemize}
\item[1.] Obtain the estimated residuals $\hat{\epsilon}_a=A-\hat{\theta}^T G$ from ordinary least squares regression of $A$ on $G$.
\item[2.] Estimate $\beta_a$  by two-stage least squares regression of $Y$ on $A$, using $\left(G-\bar{G} \right)\hat{\epsilon}_a$ as the instrument, where $\bar{G}$ is the sample mean of $G$. 
\end{itemize}
The above estimator assumes $E(A|G)=\theta^T G$, although the first step can be extended for some nonlinear, possibly unknown exposure model. Lewbel showed that $\hat{\beta}_a$ is consistent for the average causal effect which is parameterized by the scalar ${\beta}_a$  under model (\ref{eq:lewbel}) and  condition (\ref{strong IV}) as well as $\text{cov}\{G, {\epsilon}_a{\epsilon}_y\}=0,$ where ${\epsilon}_y=Y-\beta_a A$.

\subsection{Continuous exposure}

Consider the following stronger version of assumption 4   in which $\beta_a(U)=\beta_a$ with probability 1,

\begin{description}
\item[Assumption 4*:]
\item[(4a*)]
\begin{eqnarray}
E\left(  Y|A,G,U\right)     &=&\beta_{a}A + \beta_{g}(U,G)+ \beta_{u}\left(  U\right)\\
 E\left(  A|G,U\right)   &=&\alpha_{g}(U,G)  +\alpha_{u}(U).
\label{4.a1}%
\end{eqnarray}

\item[(4b*)] The orthogonality conditions
\begin{eqnarray}
&&\mbox{cov}\{\beta_{g}\left(  U,G\right),\alpha_g(U,G)|G\}=\mbox{cov}\{\beta_{g}\left(  U,G\right),\alpha_u(U)|G\}=0;\notag \\
&&\mbox{cov}\{\alpha_g(U,G),\beta_{u}\left(  U\right)|G\}=0, \notag
\end{eqnarray}
hold with probability 1.
\end{description}
Then for continuous $A$, Lemma \ref{l1} continues to hold under assumptions 1, 2 and $4^*$, where $\beta_a=\mu$ now encodes the causal effect on the outcome mean upon increasing the exposure by one unit. Condition $\left(  \ref{strong IV}\right)  $ implies that the residual error $\varepsilon_{A}=A-E(A|G)$ must be heteroscedastic, i.e.
$var(A|G)=E(\varepsilon_{A}^{2}|G)$ depends on $G.$  In the next section, we generalize this identification result in several
important directions particularly relevant to MR studies.

We note that as previously stated while $var(A|G)$ will generally depend on $G$ for binary or
discrete $A$, this may not always be the case
for continuous $A.$ However in this case, the assumption can be motivated
under an underlying model for $A$ with latent heterogeneity in the effect of
$G$ on $A.\ $Specifically, suppose that $A    =\alpha_{g}^{\ast}\left(  G,\varepsilon_{g}\right)  +U_{a}%
+\varepsilon_{a}^{\ast}$ and $E\left(  \varepsilon_{a}^{\ast}\right)    =0$,
where $\varepsilon_{g}$ and $\varepsilon_{a}^{\ast}$ are unobserved random
disturbances independent of $(G,U);$ the disturbance $\varepsilon_{g}$ may be
viewed as unobserved genetic or environmental factors independent of $G$, that
may however interact with $G$ to induce additive effect heterogeneity of G-A
associations, e.g. $\alpha^{*}_{g}(G,\epsilon_{g})=\alpha^{*}_{g} G+
\epsilon_{g} G$. Then, one can verify that the model in the above display
implies that $A=\alpha_{g}\left(  G\right)  +\varepsilon_{a}$ where
$\alpha_{g}\left(  G\right)  =E\left(  \alpha_{g}^{\ast}\left(  G,\varepsilon
_{g}\right)  |G\right)  +E\left(  U_{a}\right)  $ and
$var\left(  \varepsilon_{a}|G\right)  =var\left(  \left\{  U_{a} +\varepsilon^{*}_{a}+\alpha_{g}^{\ast}\left(  G,\varepsilon
_{g}\right)  -E\left(  \alpha_{g}^{\ast}\left(  G,\varepsilon_{g}\right)
|G\right)  \right\}  |G\right)$,
which clearly depends on $G$, provided $\alpha^{*}_{g}(g,\epsilon_{g}%
)-\alpha^{*}_{g}(0,\epsilon_{g})$ depends on $\epsilon_{g}$ for a value of
$g$, therefore implying condition $\left(  \ref{strong IV}\right)  .$ A model
for exposure which incorporates latent heterogeneity in the effects of $G$ is
quite natural in the MR context because such a model is widely considered a
leading contestant to explain the mystery of missing heritability \citep{manolio2009finding}.

Condition (\ref{strong IV}) is also related to the identification assumptions underlying an important class of bias-adjusted estimators of causal effects which leverage on gene-environment interactions when exclusion restriction of the IV is violated \citep{10.1093/ije/dyy204}. For example, in \cite{10.1093/ije/dyy204}, a genetic risk score for body mass index (BMI) is shown to interact
with a measure of social class (Townsend Deprivation Index, TDI). The genetic risk score
explains a higher proportion of variance in BMI for people with high TDI values, and therefore condition (\ref{strong IV}) holds. However, unlike \cite{10.1093/ije/dyy204}, we do not require that one observes $\epsilon_g$ in order to identify $\beta_a$, which is a key advantage.

\subsection{Identification under violation of IV Independence}

In this section, we aim to relax the IV\ independence assumption 2, by
allowing for dependence between $U$ and $G$ as displayed in Figure \ref{fig1}c.
\ Therefore, we consider replacing assumption 2 with the following weaker condition:

\begin{description}
\item[Assumption 2*.] $cov\left( \beta_u(U)%
,\alpha_u(U)|G\right) =\rho$ does not depend on $G.$
\end{description}

To illustrate assumption 2* it is instructive to consider the following
submodels of assumption 4*:
$ \beta_u(U)=\beta_{0}+\beta_{u}U$ and $\alpha_u(U)=\alpha_{0}+\alpha_{a}U$. Then assumption 2* implies
$var(U|G)=\rho/\left(  \beta_{u}\alpha_{a}\right)  $, i.e. the unmeasured
confounder $U$ has homoscedastic variance. Under assumption 2*, $E(U|G)$ is
left unrestricted therefore assumption 2 may not hold. We have the following result:
\begin{lemma}
\label{l2}
Suppose that Assumptions 1, 2*, 4* hold, then $\beta_a=\mu$ provided that
condition $\left(  \ref{strong IV}\right)  $ holds$.$
\end{lemma}

The proof of Lemma \ref{l2} is given in appendix A2. Lemma \ref{l2} implies that under
assumptions 1, 2*, 4* and condition $\left(  \ref{strong IV}\right)  ,$
$\widehat{\beta}_{a}$ continues to be consistent even if $U\notindependent G.$ As previously mentioned, MR GENIUS may be viewed as a special case of
G-estimation \citep{10.1007/978-1-4612-1842-5_4}. In fact, under assumption 4a and the
additional assumption of no unobserved confounding given $G$, i.e. if either
$U\independent A|G$ or $U\independent Y|A,G,$ the G-estimator $\widetilde{\beta}_{a}$
which solves an estimating equation of the form:
\[
0=\mathbb{P}_{n}\left[  h(G)\left\{  A-\widehat{E}(A|G)\right\}  \left\{
Y-\widetilde{\beta}_{a}A\right\}  \right]  ,
\]
is consistent and asymptotically normal for any user-specified function
$h\left(  \cdot\right)  $ (up to regularity conditions). \qquad

It is straightforward to verify that the MR GENIUS estimator $\left(
\ref{Estimator}\right)  $ solves the estimating equation:
\begin{equation}
0=\mathbb{P}_{n}\left[  \left\{  G-\mathbb{P}_{n}(G)\right\}  \left\{
A-\widehat{E}(A|G)\right\}  \left\{  Y-\widehat{\beta}_{a}A\right\}  \right]
, \label{no c}%
\end{equation}
therefore formally establishing an equivalence between MR GENIUS and
g-estimation for the choice $h(G)=G-E(G).$ Remarkably, as we have established
above, this specific choice of $h$ renders g-estimation robust to unmeasured
confounding under certain no-additive interactions conditions with unmeasured
factors used in selecting exposure levels, therefore motivating the choice of
acronym for the proposed approach.

\section{Generalizations}
\label{s:gen}

\subsection{Incorporating Covariates}

One may wish in an MR analysis to adjust for covariates, either to account for
observed confounding of the exposure effect on the outcome, or to account for
confounding of the effects of the genetic markers primarily by ancestry (known
as population stratification)\ or simply to improve efficiency. \ In order to
account for covariates $C$, we propose to solve:
\begin{equation}
0=\mathbb{P}_{n}\left[  h(C)\left\{  G-\widehat{E}(G|C)\right\}  \left\{
A-\widehat{E}(A|G,C)\right\}  \left\{  Y-\widehat{\beta}_{a}A\right\}  ,
\right]  \label{EE1}%
\end{equation}
for user-specified choice of $h,$ where $\widehat{E}(G|C)$ and $\widehat{E}%
(A|G,C)$ are consistent estimators of $E(A|G,C)$ and $\widehat{E}(G|C)$
obtained say by fitting appropriate generalized linear models. For example, as
$G$ is binary, one may specify $\mathrm{\,}$logit$\Pr(G=1|C)=\omega_{0}%
+\omega^{\prime}C$ to obtain $\widehat{E}(G|C)$ by standard likelihood
estimation of a logistic regression, and likewise when $A$ is binary, one may
obtain $\widehat{E}(A|G,C)$ by fitting a similar logistic regression, and when
$A$ is continuous, an analogous linear regression could be used instead.
Identification results established in previous Sections continue to apply by
further conditioning on $C$. 

\subsection{Incorporating Multiple IVs}

MR designs with multiple candidate genetic IVs may be used to strengthen
identification and improve efficiency. \ Multiple candidate IVs can be
incorporated by adopting a standard generalized method of moments approach$.$
Specifically, suppose that $G$ is a vector of genetic variants, we propose to obtain
$\widehat{\beta}_{a}$ by solving:%
\begin{equation}
\widehat{\beta}_{a}=\arg\min_{\beta_{a}}\mathbb{P}_{n}\left[  \widehat{U}%
^{\prime}\left(  \beta_{a}\right)  \right]  W\mathbb{P}_{n}\left[
\widehat{U}\left(  \beta_{a}\right)  \right]  \label{GMM}%
\end{equation}
where
\[
\widehat{U}\left(  \beta_{a}\right)  =\left\{  h\left(  G,C\right)
-\widehat{E}(h\left(  G,C\right)  |C)\right\}  \left\{  A-\widehat{E}%
(A|G,C)\right\}  \left\{  Y-\beta_{a}A\right\}
\]
for a user-specified function $h\left(  G,C\right)  $ of dimension $K\geq1,$
and $W$ is user-specified weight matrix. In practice, it may be convenient to
set $h\left(  G,C\right)  =G$ and $W=I_{KxK}$ the $K$ dimensional identity
matrix. Let $\overline{\beta}_{a}$ denote the corresponding estimator. A more
efficient estimator $\widehat{\beta}_{a}$ can then be obtained by solving
$\left(  \ref{GMM}\right)  $ with weight $W_{opt}=$ $\mathbb{P}_{n}\left[
\widehat{U}\left(  \overline{\beta}_{a}\right)  \widehat{U}\left(
\overline{\beta}_{a}\right)  ^{\prime}\right]  ^{-}$ where $T^{-}$ denotes the
generalized inverse of matrix $T$. \ Identification of GMM is guaranteed (at
least locally)\ provided that the second derivative with respect to $\beta_{a}$ of the GMM
objective function $\mathbb{P}_{n}\left[  \widehat{U}^{\prime}\left(
\beta_{a}\right)  \right]  W\mathbb{P}_{n}\left[  \widehat{U}\left(  \beta
_{a}\right)  \right]  $ is nonsingular at the truth, which is a generalization
of condition $\left(  \ref{strong IV}\right)  $. The asymptotic distribution
of $\widehat{\beta}_{a}$ which solves (\ref{GMM}) is described in appendix B2.

\subsection{Multiplicative causal effects}

In this Section, we consider making inferences about the multiplicative causal
effect of exposure $A,$ under the model
\begin{equation}
\frac{E\left(  Y|A=a,G,U\right)  }{E\left(  Y|A=0,G,U\right)  }=\exp\left(
\beta_{a}a\right)  , \label{5}%
\end{equation}
where for simplicity, we assume no baseline covariates, binary $A$ and scalar
$G$. Therefore, If $Y$ is binary, $\beta_{a}a$ encodes the conditional log
risk ratio $$\log\left\{  \Pr\left(  Y=1|A=a,G,U\right)/\Pr\left(
Y=1|A=0,G,U\right)  \right\},$$ which is assumed to be independent of $U$ and
$G,$ i.e. there is no multiplicative interaction between $A$ and $\left(
G,U\right).$ In order to state our identification result with an invalid IV,
consider the following assumption.

\begin{description}
\item[Assumption 4a$^{\dag}$. ]  Equations (\ref{5}) and 
\begin{eqnarray*}
E(Y|A=0,G,U)&=&\beta_{g}(U,G)+ \beta_{u}\left(  U\right)\\
E(A|G,U)&=&\alpha_{g}(U,G)+ \alpha_{u}\left(  U\right)
\end{eqnarray*}
hold.
\end{description}

\begin{lemma}
\label{l3}
Suppose that Assumptions 1, 2$^{\ast}$, 4a$^{\dag}$ and 4b* hold, then $\beta_{a}$ is the unique
solution to equation:
\begin{equation}
0=U_{\text{\it  mul}}(Y,A,G;\beta_a)\equiv E\left[  \left\{  G-E(G)\right\}  \left\{  A-E(A|G)\right\}  Y\exp\left\{
-\beta_{a}A\right\}  \right]  , \label{RR}%
\end{equation}
provided that $\partial U_{\text{\it  mul}}(Y,A,G;\beta_a)/\partial \beta_a \neq 0$ at the truth. 
\end{lemma}
The results follows upon noting that $E\left[  Y\exp\left\{  -\beta
_{a}A\right\}  |A,G,U\right]  =E\left[  Y|A=0,G,U\right].$ The proof then
proceeds as in Lemma \ref{l2}.

According to Lemma \ref{l3}, a consistent estimator of $\beta_{a}$ can be obtained
by solving an empirical version of equation $\left(  \ref{RR}\right)  $ in a
similar manner as in previous Sections. The unbiasedness property given by
equation $\left(  \ref{RR}\right)  $ continues to hold for continuous $A$
under the conditions given in Lemma \ref{l3}, and generalizations to allow for
covariates and multiple IVs can easily be deduced from previous Sections.

Interestingly, equation $\left(  \ref{RR}\right)  $ continues to hold under
case-control sampling with respect to the outcome $Y$, however note that $E(G)$ and
$E(A|G)$ must be evaluated wrt the underlying distribution for the target
population which will in general not match the corresponding distributions in
the case-control sample. To use the result\ in practice, one would either need
to obtain these quantities from an external source or one could alternatively
approximate them with the corresponding data distribution in the controls
(i.e. units with $Y=0)$ provided the outcome is sufficiently rare. In the
event sampling fractions for cases and controls are available, one could in
principle implement inverse-probability of sampling weights to consistently
estimate $E(G)$ and $E(A|G).$\ Unbiasedness under case-control sampling
follows from noting that $f\left(  A,G,U|Y=1\right)  \propto\Pr
(Y=1|A,G,U)f\left(  A,G,U\right)  ,$ and therefore 
\begin{align*}
&  E\left[  \left\{  G-E(G)\right\}  \left\{  A-E(A|G)\right\}  \exp\left\{
-\beta_{a}A\right\}  |Y=1\right] \\
&  \propto E\left[  \left\{  G-E(G)\right\}  \left\{  A-E(A|G)\right\}
\exp\left\{  -\beta_{a}A\right\}  E(Y|A,G,U)\right] \\
&  =E\left[  \left\{  G-E(G)\right\}  \left\{  A-E(A|G)\right\}  \exp\left\{
-\beta_{a}A\right\}  Y\right]  ,\\
&  =0
\end{align*}
where the last equality follows from Lemma \ref{l3}. The approach therefore extends that proposed by \cite{doi:10.1002/sim.4138} who give a detailed study of IV inferences using G-estimation under case-control sampling, in order to account for potentially invalid IVs.

\subsection{Multiplicative exposure model}

A multiplicative exposure model may also be used for count or binary exposure
under the following assumption:

\begin{description}
\item[Assumption 4$^\dag$]
\item[(4.a$^\dag$)] There is no additive $A-(U,G)$ interaction in model for $E\left(
Y|A,G,U\right)  $
\begin{equation}
E\left(  Y|A=a,G,U\right)  -E\left(  Y|A=0,G,U\right)  =\beta_{a}a
\label{4.a1}%
\end{equation}
and no additive $G-U\ $interaction in model for $E\left(  Y|A,G,U\right)  $
\begin{equation}
E\left(  Y|A=0,G=g,U\right)  -E\left(  Y|A=0,G=0,U\right)  =\beta_{g}\left(
g\right)  \label{4.a2}%
\end{equation}
for an unknown function $\beta_{g}\left(  \cdot\right)  $ that satisfies
$\beta_{g}\left(  0\right)  =0$

\item[(4.b$^\dag$)] There is no multiplicative $G-U$ interaction in model for
$E\left(  A|G,U\right)  $%
\begin{equation}
\log\frac{E\left(  A|G=g,U\right)  }{E\left(  A|G=0,U\right)  }=\alpha
_{g}\left(  g\right)
\end{equation}
for an unknown function $\alpha_{g}\left(  \cdot\right)  $ that satisfies
$\alpha_{g}\left(  0\right)  =0.$
\end{description}

MR GENIUS\ can be adapted to this setting according to the following result.
Let%
\begin{equation}
\log\frac{E\left(  A|G=g\right)  }{E\left(  A|G=0\right)  }=\varpi_{g}\left(
g\right)  ,
\end{equation}
and $U_{a}=E\left(  A|G=0,U\right)  .$

\begin{lemma}
\label{l4}
{\ Suppose that Assumptions 1, 2 and 4$^\dag$  hold, then
\[
\beta_{a}=\frac{E\left[  \left\{  G-E(G)\right\}  \left\{  A\exp(-\varpi
_{g}\left(  G\right)  )-E(A\exp\left(  -\varpi_{g}\left(  G\right)  \right)
)\right\}  Y\right]  }{E\left[  \left\{  G-E(G)\right\}  \left\{
A\exp(-\varpi_{g}\left(  G\right)  )-E(A\exp\left(  -\varpi_{g}\left(
G\right)  \right)  )\right\}  A\right]  }
%,
\]
provided that $var\left(  A|g\right)  /var\left(  A|g=0\right)  \neq
\exp\left(  \varpi_{g}\left(  g\right)  \right)  $ for at least one value of
$g.$}
\end{lemma}

The proof for Lemma \ref{l4} can be found in appendix A3. A consistent estimator of
$\beta_{a}$ is therefore obtained as in the previous section, by substituting
in consistent estimators of unknown parameters and sample averages for
expectations. To ground ideas, suppose that $\varpi_{g}\left(  g\right)
=\varpi^T_{g}g$ for vector $\varpi_{g},$ then a consistent estimator
$\widehat{\varpi}_{g}$ of $\varpi_{g}$ is given by the solution to the
estimating equation:%
\[
\mathbb{P}_{n}\left[  A\exp\left(  -\widehat{\varpi}^T_{g}G\right)  \left(
G-\mathbb{P}_{n}G\right)  \right]  =0
\]
Note that if $A$ is a rare binary exposure then $var\left(  A|g\right)
/var\left(  A|g=0\right)  \approx\exp\left(  \varpi_{g}\left(  g\right)
\right)  $ for all $g,$ therefore violating the identification condition. In
such instance, we recommend using the additive model described in the previous
section. For count data, the result rules out using a Poisson model for
exposure, however other models that accommodate over-dispersion such as the
negative binomial distribution may be used. Finally, it is straightforward to
verify that Lemma \ref{l4} continues to hold if assumption 2 is dropped to allow
for unmeasured confounding of the effects of $G$ provided that the conditional
covariance between the residual $(U_{a}/E(U_{a}|G)-1)$ and $U_{y}$ given $G$
does not depend on $G.$ Note that in this latter case $E\left(  A|G=g\right)
=\exp\left(  \varpi_{g}\left(  g\right)  \right)  =\exp\left(  \alpha
_{g}\left(  g\right)  \right)  E\left(  U_{a}|G=g\right)  $. 

\subsection{Odds ratio exposure model}

We briefly consider how MR\ GENUIS\ might be applied in a setting where
assumption $2$ is replaced by the following weaker conditional independence assumption:

\begin{description}
\item[Assumption 2$^{\dag\dag}$. ] IV conditional independence: $G\independent U|A=0$.
\end{description}

A key implication of this assumption is that the causal effect of $G$ on $Y,$
is now identified conditional on $A,$ because the assumption implies no
unmeasured confounding of the effects of $G$ on $Y.$ Note however that $G$ and
$U$ are not marginally independent. Suppose one wishes to encode the IV-exposure association on the odds ratio scale,
under the following homogeneity assumption:

\begin{description}
\item[Assumption 4a$^{\dag\dag}$] 
\item[(4a$^{\dag\dag}$)]  Equations  (\ref{5}) and $E(Y|A=0,G,U)=\beta_{g}(G)+ \beta_{u}\left(  U\right)$ hold.
\item[(4b$^{\dag\dag}$)] There is no odds ratio $G-U$ interaction in model for
$E\left(  A|G,U\right)  $%
\[
\text{logit }Pr\left(  A=1|G=g,U\right)  -\text{logit}\Pr\left(
A=1|G=0,U\right)  =\chi_{g}\left(  g\right)
\]
for an unknown function $\chi_{g}\left(  \cdot\right)  $ that satisfies
$\chi_{g}\left(  0\right)  =0.$
\end{description}

We then have the following identification result for the multiplicative causal
effect $\beta_{a}$ of model (\ref{5})

\begin{lemma}
\label{l5}
{Under assumptions 1, 2$^{^{\dag\dag}}%
$ and 4$^{\dag\dag}$,
we have that $\beta_{a}=\theta$, where $\theta$ is the unique solution to
equation:}
\[
0=E\left[  \left\{  G-E(G|A=0)\right\}  \left\{  A-E(A|G=0)\right\}
Y\exp\left\{  -\left(  \varphi_{g}\left(  G\right)  +\theta\right)  A\right\}
\right]
\]
\textit{where
\[
\varphi_{g}\left(  g\right)  =\text{logit }Pr\left(  A=1|G=g\right)
-\text{logit}\Pr\left(  A=1|G=0\right)  ,
\]
provided that $\gamma_{ag}(g)\neq0$ for some value of $g,$ with:}
\begin{align*}
\gamma_{ag}(g)= & E\left(  Y|A=1,G=g,u\right)  -E\left(  Y|A=1,G=0,u\right) \\
& -E\left(  Y|A=0,G=g,u\right)  +E\left(  Y|A=0,G=0,u\right)  \neq0.
\end{align*}
\end{lemma}
The proof for Lemma {\ref{l5}} can be found in appendix A4.
Assumption 2$^{\dag\dag}$ in fact implies that $\varphi_{g}(.) =\chi_{g}(.)$ \citep{ma2006collapsibility}. Lemma \ref{l5} establishes that under assumptions 1, 2$^{^{\dag\dag}}%
$ and 4$^{\dag\dag}$,
the multiplicative causal effect of $A$ is identified, provided that
$\gamma_{ag}(g)\neq0.$ In the proof of the Lemma we establish that under our
assumptions $\gamma_{ag}(g)=\left(  \exp\left(  \beta_{a}\right)  -1\right)
\beta_{g}\left(  g\right)  ,$ and therefore the causal effect is not
identified by the Lemma if all IVs satisfy the exclusion restriction
assumption, such that $\beta_{g}\left(  g\right)  =0$ for all $g.$ Note that
the latter assumption is empirically testable because the direct effect of $G$
on $Y$ is unconfounded. If $\beta_{g}(g) \neq0$ for some $g$, a valid test for
the causal null hypothesis can be performed by testing whether the estimating
equation given in the Lemma holds at $\theta=0$. An estimator of $\beta_{a}$
based on the Lemma is easily deduced from previous sections.

\subsection{MR GENIUS for censored failure time under a multiplicative
survival model}

Censored time-to-event endpoints are common in MR\ studies and IV methods to
address such data are increasingly of interest; recent contributions to this
literature include \cite{nie2011inference}, \cite{tchetgen2015instrumental}, \cite{li2015instrumental} and \cite{doi:10.1111/biom.12699}. While these methods have been shown to
produce a consistent causal effect estimator encoded either on the scale of
survival probabilities, or as a hazards ratio or hazards difference,
leveraging a valid IV\ which satisfies assumptions 1--3, they are not
robust to violation of any of these assumptions. In this Section, we briefly
extend MR\ GENIUS\ to survival analysis under an additive hazards model. Thus,
suppose now that $Y$ is a time-to-event outcome which satisfies the following
additive hazards model
\begin{equation}
h(y|A,U,G)=\beta_{0}\left(  y\right)  +\beta_{a}\left(  y\right)  A+\beta
_{g}(y)G+\beta_{u}\left(  y,U\right)  \label{addhaz}%
\end{equation}
where $h(y|A,U,G)$ is the hazard function of $Y$ evaluated at $y$, conditional
on $A,U$ and $G$, and the functions $\left(  \beta_{0}\left(  \cdot\right)
,\beta_{a}\left(  \cdot\right)  ,\beta_{g}(\cdot),\beta_{u}\left(  \cdot
,\cdot\right)  \right)  $ are unrestricted. The model states that conditional
on $U$, the effect of $A$ on $Y$ encoded on the additive hazards scale is
linear in $A$ for each $y$, although, the effect size $\beta_{a}\left(
y\right)  $ may vary with $y$. The model is quite flexible in the unobserved
confounder association with the outcome $\beta_{u}\left(  \cdot,\cdot\right)
$, which is allowed to remain unrestricted at each time point $y$ and across
time points. This is the model considered by \cite{tchetgen2015instrumental}
who further assumed that $\beta_{g}(y)=0$ for all $y$ by the exclusion
restriction assumption 3. Here we do not make this assumption. As usually the
case in survival analysis, $Y$ is subject to right-censoring due to drop-out,
and therefore instead of observing $Y$ for all subjects, one observes
$Y^{\ast}=\min(Y,X)$ and $\Delta=I(\min(Y,X)=Y),$ where $X$ is an independent
censoring time (i.e. independent of $Y,A,G,U).$ Let $R(y)=I(Y^{\ast}\geq y)$
denote the at-risk process and $N(y)=I(Y^{\ast}\leq y,\Delta=1)$ the counting
process associated with failure time. As discussed in \cite{doi:10.1111/biom.12699}, the additive hazards model (\ref{addhaz}) is particularly attractive
because it implies a multiplicative survival model for the joint causal effect
of $A$ and $G$ on $Y:$
\[
\frac{\Pr\left(  Y>y|A=a,G=g,U\right)  }{\Pr\left(  Y>y|A=0,G=0,U\right)
}=\exp\left\{  -\mathbf{B}_{a}\left(  y\right)  a-\mathbf{B}_{g}\left(
y\right)  g\right\}
\]
where $\mathbf{B}_{a}\left(  y\right)  =\int_{0}^{y}\beta_{a}(v)dv,\mathbf{B}%
_{g}\left(  y\right)  =\int_{0}^{y}\beta_{g}(v)dv$. Our objective is therefore
to identify and estimate $\mathbf{B}_{a}\left(  y\right)  .$   We have the
following result which extends the result of \cite{doi:10.1111/biom.12699} in order
to accommodate possible violation of the exclusion restriction assumption:

\begin{lemma}
\label{l6}
{Under assumptions 1, 2 and equations
(\ref{addhaz}) and (\ref{exp}), we have that for each $y$
\begin{equation}
0=E\left\{  W\left(  y,\mathbf{B}_{a}\left(  y\right)  ,\mathbf{B}_{g}\left(
y\right)  \right)  \right\}  , \label{iteq}%
\end{equation}
where}
\begin{align*}
W\left(  y,\mathbf{B}_{a}\left(  y\right)  ,\mathbf{B}_{g}\left(  y\right)
\right)   &  =\left[  dN(y)-d\mathbf{B}_{a}\left(  y\right)  A-d\mathbf{B}%
_{g}\left(  y\right)  G\right]\\
&\phantom{=}\times  \exp\left\{  \mathbf{B}_{a}\left(  y\right)
A+\mathbf{B}_{g}\left(  y\right)  G\right\}  R(y)h(G,A),\\
h(G,A)  &  =\left(
\begin{array}
[c]{c}%
\left(  G-E(G)\right) \\
\left(  G-E(G)\right)  \left(  A-E(A|G)\right)
\end{array}
\right)  .
\end{align*}
\end{lemma}
The proof for Lemma \ref{l6} is given in appendix A5. As in \cite{doi:10.1111/biom.12699}, the unbiasedness of equation $W\left(
y,\mathbf{B}_{a}\left(  y\right)  ,\mathbf{B}_{g}\left(  y\right)  \right)  $
suggests a way of estimating the increments $\left(  d\mathbf{B}_{a}\left(
y\right)  ,d\mathbf{B}_{g}\left(  y\right)  \right)  $ by solving an empirical
version of equation (\ref{iteq}) for each $y$ with population expectations
replaced by sample analogs, giving the following recursive estimator
\[
\left(  \widehat{\mathbf{B}}_{a}\left(  y\right)  ,\widehat{\mathbf{B}}%
_{g}\left(  y\right)  \right)  =\int_{0}^{y}\mathbb{P}_{n}\left[
\widehat{h}(A,G)^{\prime}\exp\left\{  \widehat{\mathbf{B}}_{a}\left(
s^{-}\right)  A+\widehat{\mathbf{B}}_{g}\left(  s^{-}\right)  G\right\}
dN(s)\right]  \widehat{\mathbb{M}}^{-1}\left(  s\right)  ,
\]
where $\widehat{\mathbf{B}}_{a}\left(  s^{-}\right)  $ is the value of
$\widehat{\mathbf{B}}_{a}$ right prior to $s$, and likewise for
$\widehat{\mathbf{B}}_{g}\left(  s^{-}\right)  $, and
\begin{align*}
\widehat{h}(A,G)  &  =\left(
\begin{array}
[c]{c}%
\left(  G-\widehat{E}(G)\right) \\
\left(  G-\widehat{E}(G)\right)  \left(  A-\widehat{E}(A|G)\right)
\end{array}
\right) \\
\widehat{\mathbb{M}}\left(  s\right)   &  =\mathbb{P}_{n}\left[  \left(
\begin{array}
[c]{c}%
A\\
G
\end{array}
\right)  \widehat{h}(A,G)^{\prime}R(s)\exp\left\{  \widehat{\mathbf{B}}%
_{a}\left(  s^{-}\right)  A+\widehat{\mathbf{B}}_{g}\left(  s^{-}\right)
G\right\}  \right]  .
\end{align*}
Because of its recursive structure, this estimator can be solved forward in
time starting with $\left(  d\mathbf{B}_{a}\left(  0\right)  ,d\mathbf{B}%
_{g}\left(  0\right)  \right)  =(0,0).$ The resulting estimator is a counting
process integral, therefore only changing values at observed event time. The
estimator is only defined provided $\widehat{\mathbb{M}}\left(  y\right)  $ is
invertible at each such jump time, which is essentially a necessary condition
for identification. \ The large sample behavior of the resulting estimator
follows from results derived in \cite{doi:10.1111/biom.12699} and is therefore
omitted. Note that the result relies on assumption 2 therefore ruling out
confounding of the effect of the IV\ on the outcome.

\subsection{More efficient MR GENIUS}
\label{eff}
Similar to standard g-estimation, MR GENIUS\ can be made more efficient by
incorporating information about the association between $G$ and $Y.$ This can
be achieved by the following steps:

\begin{enumerate}
\item Obtain the MR GENIUS estimator $\widehat{\beta}_{a}$ either on the
additive or multiplicative scale.

\item Define a treatment-free outcome $\widehat{Y}_{0}\left(  \widehat{\beta
}_{a}\right)  =Y-$ $\widehat{\beta}_{a}A$ 
or $\widehat{Y}_{0}\left(  \widehat{\beta}_{a}\right)  =Y\exp\left\{
-\widehat{\beta}_{a}A\right\}$.

\item Regress $\widehat{Y}_{0}\left(  \widehat{\beta}_{a}\right)  $ on $G$
using a generalized linear model with appropriate link function, and define
$\widehat{\mu}\left(  G\right)  $ a person's corresponding fitted (predicted) value.

\item Define $\widehat{\beta}_{a}^{opt}$ as the solution to
\[
0=\mathbb{P}_{n}\left[  \left\{  G-\mathbb{P}_{n}(G)\right\}  \left\{
A-\widehat{E}(A|G)\right\}  \left\{  \widehat{Y}_{0}\left(  \widehat{\beta
}_{a}^{opt}\right)  -\widehat{\mu}\left(  G\right)  \right\}  \right]
\]
with $\widehat{Y}_{0}\left(  \widehat{\beta}_{a}^{opt}\right)  =Y-$
$\widehat{\beta}_{a}^{opt}A$ or
$\widehat{Y}_{0}\left(  \widehat{\beta}_{a}^{opt}\right)  =Y\exp\left\{
-\widehat{\beta}_{a}^{opt}A\right\}$.
\end{enumerate}

If all regression models are correctly specified (including the glm for
$E(Y_{0}\left(  \beta_{a}\right)  |G)$ required in Step 3 of the above
procedure), a standard argument of semiparametric theory implies that the
asymptotic variance of $\widehat{\beta}_{a}^{opt}$ is guaranteed to be no
larger than that of $\widehat{\beta}_{a}$ \citep{10.1007/978-1-4612-1842-5_4}$.$ Interestingly, MR
GENIUS and its more efficient version\ coincide (up to asymptotic
equivalence)\ whenever nonparametric methods are used to estimate all nuisance
parameters, i..e. to estimate $E(G),$ $E(A|G)$ and $\mu\left(  G\right)
=E(Y-\beta_{a}A|G).\ $For instance, in the case of binary $G,$ such that
regression models $E(A|G)$ and $\mu\left(  G\right)  =E(Y-\beta_{a}A|G)$ are
saturated, the two estimators are exactly equal and yield identical
inferences. Both approaches also coincide if all IVs are valid, however the
above modification will tend to be more efficient with increasing number of
invalid IVs. Note that $\mu(G)$ does not necessarily have a causal
interpretation as the effect of $G$ on $Y$ may be confounded by $U$. Also note
that misspecification of a model for $\mu(G)$ does not affect consistency and
asymptotic normality of the MR GENIUS estimator of $\beta_{a}$ provided that
as we have assumed throughout, the model for $E(A|G)$ is correct.

In the case of multiplicative outcome model, it is straightforward to extend
the robustness properties of the efficient MR GENIUS\ estimator described
above under an assumption of no multiplicative interaction (rather than no
additive interaction) between $G$ and $U.$ This would simply entail replacing
$\left\{  \widehat{Y}_{0}\left(  \widehat{\beta}_{a}^{opt}\right)
-\widehat{\mu}\left(  G\right)  \right\}  $ in step 4 with $\left\{
\widehat{Y}_{0}\left(  \widehat{\beta}_{a}^{opt}\right)  \widehat{\mu}\left(
0;\widehat{\beta}_{a}^{opt}\right)  /\widehat{\mu}\left(  G;\widehat{\beta
}_{a}^{opt}\right)  \right\}  ,$ where $\widehat{\mu}\left(  G;\widehat{\beta
}_{a}^{opt}\right)  $ is the regression of $Y\exp\left\{  -\widehat{\beta}%
_{a}^{opt}A\right\}  $ on $G$ under an appropriate GLM and solving the
estimating equation in Step 4 for $\widehat{\beta}_{a}^{opt}$. One can show
using the same method of proof used throughout, that the resulting estimator
is consistent for the causal effect of interest under violation of both
assumptions 2 and 3, under an assumption analogous to assumption 2*. Note
however that $\widehat{\mu}\left(  g\right)  /\widehat{\mu}\left(  0\right)  $
would now need to be consistent for $E(Y|A=0,G=g)/E(Y|A=0,G=0).$ It is
likewise possible to modify the above procedure to accommodate a
multiplicative exposure model by substituting in $\left\{  A\exp
(-\widehat{\varpi}_{g}\left(  G\right)  )-\mathbb{P}_{n}(A\exp\left(
-\widehat{\varpi}_{g}\left(  G\right)  \right)  )\right\}  $ for $\left\{
A-\widehat{E}(A|G)\right\}  $ in Step 4.

\section{Simulation Study}
\label{s:sim}
\subsection{Single IV}

We investigate the finite-sample properties of MR GENIUS proposed above and
compare them with existing estimators under a variety of settings. For a
single binary IV $G$, we generate independent and identically distributed
$(G_{i},U_{i},A_{i},Y_{i})$, $i=1,2,...,n$ as follows:
\begin{eqnarray*}
G_{i}  &  \sim\mbox{Bernoulli}(p=0.5),\quad Y_{i}   \sim\mbox{N}(\alpha G_{i}+\beta A_{i}+U_{i},1^{2}),
\end{eqnarray*}
where for binary exposure $A$,
\begin{eqnarray*}
\epsilon_{i}  &  \sim&\mbox{truncated N}(a=0.2,b=0.5,\mu=0.35,\sigma^{2}%
=1^{2}),\quad U_{i}    =\phi_{b}G_{i}+\epsilon_{i},\\
A_{i}  &  \sim&\mbox{Bernoulli}\left(  p_{i}=\frac{\exp{(\gamma_{b}G_{i})}%
}{1+\exp{(\gamma_{b}G_{i})}}+U_{i}-E(U_{i}|G_{i})\right)  ,
\end{eqnarray*}
where $\epsilon_{i}$ is appropriately bounded to ensure that $p$ falls in the
unit interval, and for continuous $A$, 
\begin{eqnarray*}
U_{i}   =\phi_{c}G_{i}+N(0,1^{2}),\quad A_{i}    \sim\mbox{N}\left(  \gamma_{c}G_{i}+U_{i},|\lambda_{0}+\lambda
_{1}G_{i}|^{2}\right).
\end{eqnarray*}
The data generating mechanism satisfies assumptions 2* and 4*. We set
$\gamma_{b}=-0.5$ or $-1$ (binary $A$), and $\gamma_{c}=-1$, $\lambda_{0}=1$,
$\lambda_{1}=1$ or $5$ (continuous $A$) which satisfy both Assumption 1 and
condition $(5)$. Assumptions 2 and 3 are violated when we set $\phi_{b}=-0.2$,
$\phi_{c}=-2$ and $\alpha=-0.5$ respectively. The causal parameter is set
equal to $\beta=0.5$ throughout this simulation. The IV strength is tuned by
varying the values of $\gamma_{b}$ and $\lambda_{1}$, for binary and
continuous $A$ respectively.

MR GENIUS is implemented as given in \ref{Estimator}, with $\hat{E}(A|G)$ estimated with
linear or logistic regression when $A$ is continuous or binary, respectively.
In this single-IV setting, we also implement the two-stage least squares
(TSLS) estimator, which is the most common approach used in practice. The
simulation results based on 1000 replicates at sample sizes $n=500$ and
$n=1000$ are summarized in Tables \ref{tab:single_cont} and \ref{tab:single_binary}, for continuous and binary exposure
respectively. When Assumptions 2 and 3 both hold, TSLS and MR GENIUS have
small bias regardless of sample size.  When the IV is
invalid, TSLS is biased while in accordance with
theory MR GENIUS continues to have small bias.

\begin{table}

\caption{Monte Carlo results of MR GENIUS and TSLS estimation of $\beta
_{0}=0.5$ with continuous exposure and single IV at two different strengths
($\lambda_{1}=1,5$). The first and second rows' results for each estimator
correspond to sample sizes $n=500,1000$ respectively.}%
\label{tab:single_cont}
\begin{center}

\begin{tabular}
[c]{rcccccc}%
\hline & \multicolumn{3}{c}{$|\lambda_{1}|=1$} &
\multicolumn{3}{c}{$|\lambda_{1}|=5$}\\
& \multicolumn{1}{c}{\texttt{TTT$^{\dagger}$}} &
\multicolumn{1}{c}{\texttt{TTF}} & \multicolumn{1}{c}{\texttt{TFF}} &
\multicolumn{1}{c}{\texttt{TTT}} & \multicolumn{1}{c}{\texttt{TTF}} &
\multicolumn{1}{c}{\texttt{TFF}}\\
\hline

\multicolumn{6}{l}{Median absolute value of bias} & \\[3pt]%
\quad\quad\quad\quad MR GENIUS & 0.00 & 0.00 & 0.00 & 0.00 & 0.00 & 0.00\\
& 0.00 & 0.00 & 0.00 & 0.00 & 0.00 & 0.00\\
TSLS & 0.00 & 0.50 & 0.83 & 0.00 & 0.51 & 0.84\\
& 0.00 & 0.50 & 0.83 & 0.00 & 0.50 & 0.83\\[12pt]%
\multicolumn{6}{l}{Monte Carlo SD$^{\ddagger}$} & \\[3pt]%
MR GENIUS & 0.08 & 0.08 & 0.08 & 0.02 & 0.02 & 0.02\\
& 0.06 & 0.06 & 0.06 & 0.01 & 0.01 & 0.01\\
TSLS & 0.12 & 0.13 & 0.05 & 0.13 & 0.25 & 0.11\\
& 0.09 & 0.09 & 0.04 & 0.09 & 0.15 & 0.08\\[12pt]%
\hline
\\
\end{tabular}
\end{center}
\par
{\footnotesize {$^{\dagger}$: \texttt{TTT}: IV assumptions 1--3
hold; \texttt{TTF}: IV assumption 3 does not hold;
\texttt{TFF}: both IV assumptions 2 and 3 do not
hold.\newline$^{\ddagger}$: Robust normal-consistent estimate obtained from
dividing the interquartile range of causal effect estimates by 1.349.\newline}
}
\end{table}

\begin{table}
\caption{Monte Carlo results of MR GENIUS and TSLS estimation of $\beta
_{0}=0.5$ with binary exposure and single IV at two different strengths
($\gamma_{b}=-0.5,-1$). The first and second rows' results for each estimator
correspond to sample sizes $n=500,1000$ respectively.}%
\label{tab:single_binary}
\begin{center}

\begin{tabular}
[c]{rcccccc}%
\hline & \multicolumn{3}{c}{$|\gamma_{b}|=0.5$} &
\multicolumn{3}{c}{$|\gamma_{b}|=1$}\\
& \multicolumn{1}{c}{\texttt{TTT$^{\dagger}$}} &
\multicolumn{1}{c}{\texttt{TTF}} & \multicolumn{1}{c}{\texttt{TFF}} &
\multicolumn{1}{c}{\texttt{TTT}} & \multicolumn{1}{c}{\texttt{TTF}} &
\multicolumn{1}{c}{\texttt{TFF}}\\
\hline
\multicolumn{6}{l}{Median absolute value of bias} & \\[3pt]%
\quad\quad\quad\quad MR GENIUS & 0.01 & 0.01 & 0.01 & 0.03 & 0.03 & 0.01\\
& 0.01 & 0.01 & 0.00 & 0.01 & 0.01 & 0.00\\
TSLS & 0.00 & 4.05 & 2.16 & 0.00 & 2.17 & 1.61\\
& 0.02 & 4.07 & 2.18 & 0.01 & 2.18 & 1.61\\[12pt]%
\multicolumn{6}{l}{Monte Carlo SD$^{\ddagger}$} & \\[3pt]%
MR GENIUS & 0.54 & 0.54 & 0.32 & 0.35 & 0.35 & 0.34\\
& 0.39 & 0.39 & 0.22 & 0.25 & 0.25 & 0.24\\
TSLS & 0.79 & 1.37 & 0.37 & 0.41 & 0.52 & 0.27\\
& 0.55 & 1.06 & 0.28 & 0.29 & 0.37 & 0.18\\[12pt]%
\hline 
\\
\end{tabular}
\end{center}
\par
{\footnotesize {$^{\dagger}$: \texttt{TTT}: IV assumptions 1--3
hold; \texttt{TTF}: IV assumption 3 does not hold;
\texttt{TFF}: both IV assumptions 2 and 3 do not
hold.\newline$^{\ddagger}$: Robust normal-consistent estimate obtained from
dividing the interquartile range of causal effect estimates by 1.349.\newline}
}
\end{table}
\subsection{Multiple IVs}

Here we generate i.i.d. $L_{i}=({G}_{i},U_{i},A_{i},Y_{i})$, $i=1,2,...,n$,
with $p_{G}=10$ IVs from:
\begin{eqnarray*}
{G}_{ij}   \sim\mbox{Bernoulli}(p=0.5),j=1,2,...,p_{G}, \quad Y_{i}    \sim\mbox{N}({\alpha}^{T}{G}_{i}+\beta A_{i}+U_{i},1^{2}),
\end{eqnarray*}
where $G_{i}=(G_{i1},G_{i2},...,G_{ip_{G}})^{T}$. For binary exposure,
\begin{eqnarray*}
\epsilon_{i}  &  \sim&\mbox{Truncated N}(a=0.2,b=0.5,\mu=0.35,\sigma^{2}%
=1^{2}),\quad U_{i}    ={\phi_{b}}^{T}{G}_{i}+\epsilon_{i},\\
A_{i}  &  \sim&\mbox{Bernoulli}\left(  p_{i}=\frac{\exp{({\gamma_{b}}^{T}%
{G}_{i})}}{1+\exp{({\gamma_{b}}^{T}{G}_{i})}}+\left[  U_{i}-E(U|{G}%
_{i})\right]  \right)  ,
\end{eqnarray*}
where $\epsilon_{i}$ is appropriately bounded to ensure that $p_{i}$ falls in
the unit interval, and for continuous exposure, 
\begin{eqnarray*}
U_{i}    ={\phi_{c}}^{T}{G}_{i}+\mbox{N}(0,1^{2}),\quad
A_{i}    \sim\mbox{N}\left(  {\gamma_{c}}^{T}{G}_{i}+U_{i},|\lambda
_{0}+{\lambda}_{1}^{T}{G}_{i}|^{2}\right).
\end{eqnarray*}
For binary exposure, $\gamma_{b}\sim\mbox{Uniform}(-0.15,-0.05)$ so that IV strength is variable, while in the continuous exposure case  $\gamma_{c}\sim\mbox{Uniform}(-3,-2)$
and $(\lambda_{0},\lambda_{1})$ is set to $(1,0.5)$. We first
generate an ideal scenario in which all 10 IVs are valid and satisfy
assumptions 1--3, next we consider scenarios where the first three, six or all
of the IVs are invalid. With three invalid IVs, $\alpha^{T}=-0.5\cdot
(1,1,1,0,...,0)$ and $\phi_{c}^{T}=-0.25\cdot(1,1,1,0,...,0),\phi_{b}%
^{T}=-0.05\cdot(1,1,1,0,...,0)$ when assumption 3 or 2 is violated,
respectively; with six invalid IVs, $\alpha^{T}=-0.25(1,1,2,2,4,4,0,...,0)$
and $\phi_{c}^{T}=-0.25\cdot(0.5,0.5,1,1,2,2,0,...,0),\phi_{b}^{T}%
=-0.01\cdot(1,1,3,3,5,5,0,...,0)$ accordingly. When all IVs are invalid,
$\alpha\sim\mbox{Uniform}(-2,-0.5)$, $\phi_{c}\sim\mbox{Uniform}(-2,-0.5)$ and $\phi
_{b}\sim\mbox{Uniform}(-0.02,-0.01)$. The setting with three invalid IVs
investigates the condition in which fewer than 50\% of the IVs are invalid
\citep{doi:10.1080/01621459.2014.994705, doi:10.1080/01621459.2018.1498346}; in the setting with six invalid
IVs this condition is violated, but the set of valid IVs form the largest
group according to the plurality rule \citep{doi:10.1111/rssb.12275}.

MR GENIUS is implemented as the solution to (\ref{GMM}) with optimal weight;
a more efficient version of MR GENIUS as described in section \ref{eff} is also
implemented. MR-Egger, TSLS and sisVIVE are implemented using the R packages \texttt{MendelianRandomization},
\texttt{AER} and \texttt{sisVIVE} \citep{mr, kleiber2008applied, sisvive} respectively, under default settings. The adaptive Lasso and TSHT estimation methods are
implemented as described in \cite{doi:10.1080/01621459.2018.1498346} and \cite{doi:10.1111/rssb.12275}
respectively. We also implement post-adaptive Lasso which uses adaptive Lasso
for the purpose of selecting valid IVs but not in the process of estimating
the causal effect. We also implement the oracle TSLS which assumes the set of
valid IVs to be known a priori.

Simulation results based on 1000 replications for sample sizes of $n=1000$ and
$2000$ with continuous exposure are presented in Table \ref{tab:multiple_cont_1}.
When there are zero or three invalid IVs (majority rule holds), the sisVIVE,
adaptive, post-adaptive Lasso and TSHT estimators exhibit small bias which
becomes negligible at sample size of $n=2000$. Adaptive Lasso and TSHT on average correctly identifies invalid IVs,
while sisVIVE on average selects four IVs as invalid when there are three in
truth (see Table \ref{tab:selected} for results on IV selection). The naive TSLS estimator
performs well in terms of bias only when all IVs are valid; as
expected, it is biased in all other
settings with at least one invalid IV. Post-adaptive Lasso is generally less
biased in finite sample than adaptive Lasso. Post-adaptive Lasso and oracle
TSLS perform similarly in terms of bias and efficiency when
the majority rule holds, in agreement with theory since they are
asymptotically equivalent under these settings \citep{doi:10.1080/01621459.2018.1498346}. MR GENIUS also has small bias
at all sample sizes and its bias becomes negligible at $n=2000$. When six IVs are invalid and the majority rule is violated, sisVIVE and
adaptive/post-adaptive Lasso are significantly biased, with no improvement as
sample size increases. On average, sisVIVE and
adaptive Lasso select 7 to 8 IVs as invalid when only six are actually invalid,
and fails to select all the IVs as invalid when all in fact are. TSHT is also biased when all IVs are invalid (with about 5 of the IVs
selected as invalid on average in this case); however when six IVs are
invalid, the plurality rule holds and its bias diminishes at $n=2000$.
The efficiency of all estimators
generally decreases with increasing number of invalid IVs. 

The bias of MR GENIUS improves with increasing sample size
when six or all IVs are invalid.
The efficient MR GENIUS is generally less biased and more
efficient compared to MR GENIUS, especially when more IVs are invalid. MR-Egger shows little bias when any or all of the IVs are invalid, provided only assumption 3 (exclusion restriction) is violated, in agreement with theory.  However, MR-Egger is generally more
biased when the invalid IVs violate both
assumptions 2 and 3, which corresponds to a violation of the InSIDE
assumption \citep{bowden2015mendelian}. 

Simulation results with a binary exposure are reported in Table \ref{tab:multiple_bin_1}; the
conclusions are mostly qualitatively similar to those in the continuous
exposure setting. However, when there are six invalid IVs, TSHT is biased with no improvement as sample size increases. While
the exposure is generated under a logit model (upon marginalizing over $U)$,
TSHT assumes a linear model which is misspecified in this simulation study. In
addition, because the exposure is binary, most if not all IVs are weakly
associated with $A$ on the additive scale. Weak IVs may not be selected as
valid IVs in the first thresholding step of TSHT (the number of IVs selected
as relevant is 3.2 on average at $n=2000$); even if they are included,
their inclusion may lead to incorrect inference in the subsequent estimation
step (the number of IVs selected as relevant but invalid is close to 0.5 on
average at sample size of $n=2000$, when in fact 6 are invalid). MR-Egger also appears to exhibit more bias, since the exposure model is misspecified as the linear probability model.

\begin{table}[!htbp]
\caption{{\protect\footnotesize Simulation results for estimation of $\beta_{0}=0.5$ with continuous exposure and
$p_{G}=10$ IVs. The two
rows of results for each estimator correspond to sample sizes of $n=1000$ and
$n=2000$ respectively.}}%
\label{tab:multiple_cont_1}
\begin{center}
 \resizebox{0.6\textheight}{!}{		\begin{tabular}{rrrrrrrrrrrrrrrr rrrrrrrrrrrrrrrr}
			\hline
			& & \multicolumn{3}{c}{\tt TTF$^\dagger$}& \multicolumn{3}{c}{\tt TFF} & & & \multicolumn{3}{c}{\tt TTF}& \multicolumn{3}{c}{\tt TFF} \\
		\multicolumn{1}{c}{\#invalid IV} & 0 &3 & 6 & 10 & 3 & 6& 10  & & 0 &3 & 6 & 10 & 3 & 6& 10 \\
			\hline
			& \multicolumn{7}{c}{Median absolute value of bias}	 &  &\multicolumn{7}{c}{Monte Carlo SD$^\ddagger$}	\\[3pt]
			\quad \quad 	\quad \quad	MR GENIUS &$0.01$ & $0.01$ & $0.02$ & $0.05$ & $0.02$ & $0.03$ & $0.12$& &$0.03$ & $0.03$ & $0.04$ & $0.10$ & $0.04$ & $0.05$ & $0.10$ \\ 
&$0.00$ & $0.00$ & $0.01$ & $0.02$ & $0.01$ & $0.02$ & $0.06$&&$0.02$ & $0.02$ & $0.02$ & $0.04$ & $0.03$ & $0.03$ & $0.10$ \\     [12pt]
Efficient MR GENIUS&$0.01$ & $0.01$ & $0.01$ & $0.01$ & $0.01$ & $0.01$ & $0.03$&& $0.04$ & $0.04$ & $0.03$ & $0.03$ & $0.04$ & $0.04$ & $0.04$  \\ 
&$0.00$ & $0.00$ & $0.00$ & $0.01$ & $0.00$ & $0.00$ & $0.01$&& 
$0.02$ & $0.02$ & $0.02$ & $0.02$ & $0.02$ & $0.02$ & $0.03$  \\  [12pt]

			TSLS&$0.00$ & $0.06$ & $0.13$ & $0.48$ & $0.12$ & $0.26$ & $0.66$&&$0.01$ & $0.01$ & $0.01$ & $0.10$ & $0.01$ & $0.02$ & $0.04$\\ 
&$0.00$ & $0.06$ & $0.13$ & $0.48$ & $0.12$ & $0.26$ & $0.65$ && $0.01$ & $0.01$ & $0.01$ & $0.10$ & $0.01$ & $0.01$ & $0.04$\\   [12pt]

			Oracle TSLS&$-$ & $0.00$ & $0.00$ & $-$ & $0.00$ & $0.00$ & $-$ &&
$-$ & $0.01$ & $0.02$ & $-$ & $0.01$ & $0.02$ & $-$ \\ 
&$-$ & $0.00$ & $0.00$ & $-$ & $0.00$ & $0.00$ & $-$ && $-$ & $0.01$ & $0.01$ & $-$ & $0.01$ & $0.01$ & $-$\\   [12pt]

sisVIVE&$0.00$ & $0.03$ & $0.11$ & $0.48$ & $0.04$ & $0.22$ & $0.65$  &&$0.01$ & $0.01$ & $0.02$ & $0.10$ & $0.01$ & $0.03$ & $0.10$ \\ 
&$0.00$ & $0.02$ & $0.11$ & $0.48$ & $0.02$ & $0.22$ & $0.65$ &&$0.01$ & $0.01$ & $0.02$ & $0.10$ & $0.01$ & $0.02$ & $0.10$ \\  [12pt]

			ALasso&$0.00$ & $0.02$ & $0.09$ & $0.48$ & $0.03$ & $0.21$ & $0.65$ &&$0.01$ & $0.01$ & $0.03$ & $0.10$ & $0.01$ & $0.04$ & $0.10$\\ 
&$0.00$ & $0.02$ & $0.10$ & $0.48$ & $0.02$ & $0.20$ & $0.65$&&$0.01$ & $0.01$ & $0.03$ & $0.10$ & $0.01$ & $0.03$ & $0.10$   \\ [12pt]

			post-ALasso&$0.00$ & $0.00$ & $0.10$ & $0.48$ & $0.00$ & $0.20$ & $0.65$ &&$0.01$ & $0.02$ & $0.05$ & $0.10$ & $0.01$ & $0.05$ & $0.10$\\ 
&$0.00$ & $0.00$ & $0.11$ & $0.48$ & $0.00$ & $0.20$ & $0.65$ &&$0.01$ & $0.01$ & $0.04$ & $0.10$ & $0.01$ & $0.02$ & $0.10$\\ [12pt]

			TSHT&$0.00$ & $0.01$ & $0.06$ & $0.48$ & $0.00$ & $0.11$ & $0.65$ &&$0.01$ & $0.02$ & $0.03$ & $0.10$ & $0.01$ & $0.10$ & $0.10$\\ 
&$0.00$ & $0.00$ & $0.06$ & $0.49$ & $0.00$ & $0.04$ & $0.66$ & &$0.01$ & $0.01$ & $0.03$ & $0.10$ & $0.01$ & $0.10$ & $0.10$\\  [12pt]

			MR-Egger&$0.02$ & $0.01$ & $0.02$ & $0.11$ & $0.41$ & $0.82$ & $0.54$&&$0.10$ & $0.10$ & $0.20$ & $0.30$ & $0.20$ & $0.20$ & $0.30$ \\ 
&$0.01$ & $0.02$ & $0.02$ & $0.08$ & $0.46$ & $0.93$ & $0.54$&& $0.05$ & $0.10$ & $0.20$ & $0.40$ & $0.10$ & $0.20$ & $0.30$ \\ [12pt]

\hline
		\end{tabular}
		}
\end{center}
\par
{\footnotesize {$^{\dagger}$: For the invalid IVs, \texttt{TTF}: IV assumption
3 does not hold; \texttt{TFF}: both IV assumptions
2 and 3 do not hold.\newline$^{\ddagger}$: Robust
normal-consistent estimate obtained from dividing the interquartile range of
causal effect estimates by 1.349.\newline} }
\end{table}

\begin{table}[!htbp]
\caption{{\protect\footnotesize Simulation results for estimation of $\beta_{0}=0.5$ with binary exposure and
$p_{G}=10$ IVs. The two
rows of results for each estimator correspond to sample sizes of $n=1000$ and
$n=2000$ respectively.}}%
\label{tab:multiple_bin_1}
\begin{center}
\resizebox{0.6\textheight}{!}{		\begin{tabular}{rrrrrrrrrrrrrrrr rrrrrrrrrrrrrrrr}
			\hline
			& & \multicolumn{3}{c}{\tt TTF$^\dagger$}& \multicolumn{3}{c}{\tt TFF} & & & \multicolumn{3}{c}{\tt TTF}& \multicolumn{3}{c}{\tt TFF} \\
		\multicolumn{1}{c}{\#invalid IV} & 0 &3 & 6 & 10 & 3 & 6& 10  & & 0 &3 & 6 & 10 & 3 & 6& 10 \\
			\hline
			& \multicolumn{7}{c}{Median absolute value of bias}	 &  &\multicolumn{7}{c}{Monte Carlo SD$^\ddagger$}	\\[3pt]
			\quad \quad 	\quad \quad	MR GENIUS &
$0.07$ & $0.08$ & $0.23$ & $0.63$ & $0.08$ & $0.25$ & $0.69$& &$0.91$ & $1.00$ & $1.24$ & $2.23$ & $1.01$ & $1.26$ & $2.21$ \\ 
&$0.00$ & $0.08$ & $0.20$ & $0.66$ & $0.09$ & $0.21$ & $0.66$&&$0.85$ & $0.93$ & $1.14$ & $2.22$ & $0.94$ & $1.17$ & $2.10$ \\     [12pt]
Efficient MR GENIUS&$0.02$ & $0.03$ & $0.05$ & $0.12$ & $0.04$ & $0.05$ & $0.09$&&$0.90$ & $0.91$ & $0.93$ & $1.01$ & $0.91$ & $0.93$ & $0.99$ \\ 
&$0.02$ & $0.01$ & $0.00$ & $0.07$ & $0.01$ & $0.01$ & $0.04$&& 
$0.82$ & $0.82$ & $0.82$ & $0.86$ & $0.82$ & $0.82$ & $0.81$  \\  [12pt]

			TSLS&$0.04$ & $1.70$ & $5.88$ & $20.24$ & $1.87$ & $6.20$ & $20.38$ &&$0.51$ & $1.82$ & $2.94$ & $7.35$ & $1.98$ & $3.09$ & $7.24$\\ 
&$0.02$ & $2.31$ & $8.32$ & $28.53$ & $2.53$ & $8.74$ & $28.22$  && $0.44$ & $1.64$ & $2.79$ & $7.53$ & $1.83$ & $2.95$ & $7.61$\\   [12pt]

			Oracle TSLS&$-$ & $0.05$ & $0.09$ & $-$ & $0.05$ & $0.09$ & $-$  &&
$-$ & $0.59$ & $0.83$ & $-$ & $0.59$ & $0.83$ & $-$ \\ 
&$-$ & $0.03$ & $0.04$ & $-$ & $0.03$ & $0.04$ & $-$  && $-$ & $0.51$ & $0.66$ & $-$ & $0.51$ & $0.66$ & $-$\\   [12pt]

sisVIVE&$0.04$ & $0.53$ & $3.80$ & $20.12$ & $0.52$ & $3.93$ & $20.26$ &&$0.51$ & $0.78$ & $3.44$ & $7.31$ & $0.77$ & $3.65$ & $7.28$ \\ 
&$0.02$ & $0.45$ & $5.70$ & $28.21$ & $0.45$ & $5.95$ & $28.08$ &&$0.44$ & $0.59$ & $4.13$ & $7.76$ & $0.59$ & $4.38$ & $7.93$\\  [12pt]

			ALasso&$0.04$ & $0.37$ & $2.04$ & $20.09$ & $0.36$ & $2.04$ & $20.29$  &&$0.51$ & $0.62$ & $2.94$ & $7.35$ & $0.60$ & $3.13$ & $7.17$\\ 
&$0.02$ & $0.31$ & $3.46$ & $28.27$ & $0.30$ & $3.59$ & $28.00$&&$0.42$ & $0.46$ & $4.08$ & $7.48$ & $0.46$ & $4.34$ & $7.85$  \\ [12pt]

			post-ALasso&$0.04$ & $0.10$ & $1.60$ & $20.02$ & $0.09$ & $1.58$ & $19.95$&&$0.50$ & $0.62$ & $2.68$ & $7.50$ & $0.60$ & $2.82$ & $7.12$ \\ 
&$0.02$ & $0.02$ & $2.79$ & $28.15$ & $0.02$ & $2.78$ & $27.72$ &&$0.43$ & $0.50$ & $3.77$ & $7.57$ & $0.50$ & $3.91$ & $8.06$ \\ [12pt]

			TSHT&$0.03$ & $0.27$ & $3.68$ & $17.80$ & $0.24$ & $3.76$ & $17.50$ &&$0.65$ & $1.46$ & $5.58$ & $8.88$ & $1.35$ & $5.93$ & $7.96$ \\ 
&$0.00$ & $0.07$ & $3.35$ & $22.41$ & $0.06$ & $3.57$ & $22.05$ & &$0.56$ & $0.70$ & $7.21$ & $9.62$ & $0.70$ & $7.57$ & $9.44$\\  [12pt]

			MR-Egger&$0.01$ & $1.21$ & $4.80$ & $15.05$ & $1.34$ & $5.06$ & $16.09$&&$0.93$ & $4.38$ & $7.64$ & $15.34$ & $4.80$ & $8.08$ & $16.19$\\ 
&$0.04$ & $0.32$ & $5.69$ & $16.04$ & $0.35$ & $6.01$ & $16.06$ && $0.82$ & $5.06$ & $9.24$ & $16.76$ & $5.61$ & $9.72$ & $19.14$  \\ [12pt]

							\hline
		\end{tabular}
		}
\end{center}
\par
{\footnotesize {$^{\dagger}$: For the invalid IVs, \texttt{TTF}: IV assumption
3 does not hold; \texttt{TFF}: both IV assumptions
2 and 3 do not hold.\newline$^{\ddagger}$: Robust
normal-consistent estimate obtained from dividing the interquartile range of
causal effect estimates by 1.349.\newline} }
\end{table}

\begin{table}[!htbp]
\caption{{\protect\footnotesize Average number of IVs selected as invalid by
adaptive Lasso and sisVIVE, and average number of IVs selected as relevant
($\hat{S}$) and relevant but invalid ($\hat{I}$) by TSHT. The two rows of
results for each estimator correspond to sample sizes of $n=1000$ and $n=2000$  respectively.}}%
\label{tab:selected}
\begin{center}
 \resizebox{0.6\textheight}{!}{		\begin{tabular}{rrrrrrrrrrrrrrrr}
			\hline
			& & \multicolumn{3}{c}{\tt TTF$^\dagger$}& \multicolumn{3}{c}{\tt TFF} & & & \multicolumn{3}{c}{\tt TTF$^\dagger$}& \multicolumn{3}{c}{\tt TFF} \\
		\multicolumn{1}{c}{\#invalid IV} & 0 &3 & 6 & 10 & 3 & 6& 10 & & 0 &3 & 6 & 10 & 3 & 6& 10  \\
			\hline
			&\multicolumn{6}{c}{Continuous exposure}	 & & &\multicolumn{7}{c}{Binary exposure}	 \\[3pt]
			\quad \quad 	\quad \quad	ALasso &$0.0$ & $3.1$ & $5.5$ & $4.8$ & $3.0$ & $7.3$ & $4.4$ &&$0.1$ & $3.2$ & $5.0$ & $1.0$ & $3.1$ & $5.1$ & $0.9$ \\
&$0.0$ & $3.0$ & $6.8$ & $5.9$ & $3.0$ & $7.8$ & $5.5$ &&$0.1$ & $3.1$ & $5.0$ & $1.0$ & $3.1$ & $5.1$ & $0.9$\\
			sisVIVE & $0.0$ & $3.8$ & $7.1$ & $6.2$ & $4.2$ & $7.9$ & $5.6$  && $0.0$ & $3.8$ & $5.0$ & $1.1$ & $3.8$ & $5.0$ & $1.1$\\
&$0.0$ & $3.9$ & $7.8$ & $7.3$ & $4.2$ & $8.3$ & $6.8$  &&$0.0$ & $3.7$ & $5.1$ & $1.3$ & $3.7$ & $5.2$ & $1.4$ \\

			TSHT ($\hat{I}$) & $0.0$ & $2.4$ & $4.5$ & $3.4$ & $3.0$ & $6.8$ & $2.7$   &&$0.0$ & $0.7$ & $0.4$ & $0.0$ & $0.7$ & $0.5$ & $0.1$ \\
&$0.0$ & $3.0$ & $7.1$ & $5.7$ & $3.0$ & $5.6$ & $5.1$  &&$0.0$ & $0.6$ & $0.5$ & $0.1$ & $0.6$ & $0.5$ & $0.1$  \\

			TSHT ($\hat{S}$)& $10.0$ & $10.0$ & $10.0$ & $10.0$ & $10.0$ & $10.0$ & $10.0$ && $4.7$ & $4.7$ & $4.7$ & $4.7$ & $4.7$ & $4.7$ & $4.5$\\
&$10.0$ & $10.0$ & $10.0$ & $10.0$ & $10.0$ & $10.0$ & $10.0$ &&$3.2$ & $3.2$ & $3.2$ & $3.2$ & $3.2$ & $3.2$ & $3.4$  \\

							\hline
		\end{tabular}
		}
\end{center}
\par
{\footnotesize {$^{\dagger}$: For the invalid IVs, \texttt{TTF}: IV assumption
3 does not hold; \texttt{TFF}: both IV assumptions
2 and 3 do not hold.\newline} }
\end{table}

\section{Data Application}
\label{s:app}
The prevalence of type 2 diabetes mellitus is increasing across all age groups
in the United States possibly as a consequence of the obesity epidemic. Many
epidemiological studies have suggested that individuals with type 2 diabetes
mellitus (T2D) are at higher risk of various memory impairments which are
highly associated with dementia and Alzheimer's Disease. However, such
observational studies are well known to be vulnerable to confounding bias.
Therefore, obtaining an unbiased estimate of the association between diabetes
status and cognitive functioning is key to predicting the future health burden
in the population and to evaluating the effectiveness of possible public
health interventions.

In order to illustrate the proposed MR approach, we used data from the Health
and Retirement Study, a cohort initiated in 1992 with repeated assessments
every 2 years. We used externally validated genetic predictors of type 2
diabetes as IVs to estimate effects on memory functioning among HRS
participants. The Health and Retirement Study is a well-documented nationally
representative sample of persons aged 50 years or older and their spouses
\citep{10.2307/146277}. Genotype data were collected on a subset of
respondents in 2006 and 2008. Genotyping was completed on the Illumina
Omni-2.5 chip platform and imputed using the 1000G phase 1 reference panel and
filed with the Database for Genotypes and Phenotypes (dbGaP, study accession
number: phs000428.v1.p1) in April 2012. Exact information on the process
performed for quality control is available via Health and Retirement Study and
dbGaP21 \citep{mailman2007ncbi}. From the 12,123 participants for whom genotype data
was available, we restricted the sample to 7,738 non-hispanic white persons
with valid self-reported diabetes status at baseline and memory assessment
score two years later. Self-reported diabetes in the Health and Retirement
Study has been shown to have 87\% sensitivity and 97\% specificity for
Hemoglobin A1c defined diabetes among non-Hispanic white HRS participants
\citep{white}. Memory was assessed by immediate and delayed recall of a
10-word list plus the proxy assessments for severely impaired individuals. The
validity and reliability of these measures have been documented elsewhere
\citep{5620,wu}.

Standard MR relies on the assumption that all 39 SNPs affect a person's memory
score at follow-up only through baseline diabetes status which is unlikely,
even if all 39 SNPs only affect memory through diabetes. This is because there
is likely to be a nonnegligible direct effect from one of the SNPs to diabetes
incidence among persons who are diabetes-free at baseline. This would
constitute a violation of the exclusion restriction and therefore would
invalidate a standard MR analysis\ for assessing effects of baseline diabetes
on memory score at follow-up. Nonetheless, although possibly positively biased
under the alternative hypothesis, the two-stage regression estimator could
still be interpreted as a valid test of the null hypothesis of no association
between diabetes disease (whether baseline or time-updated) and memory score.
It may also be true that unknown pleiotropic effects of at least one of the
SNPs exists through a pathway not involving diabetes, which would constitute
an even more serious violation, as it would also invalidate our MR\ analysis
as a valid test of a causal association between diabetes and memory
functioning. In light of these possible limitations a more robust MR analysis
is naturally of interest.

We used GENIUS to estimate the relationship between diabetes status (coded $1$
for diabetic and $0$ otherwise) and memory score. As genetic instruments, we
used 39 independent single nucleotide polymorphisms previously established to
be significantly associated with diabetes \citep{morris2012large}.

We first performed an observational analysis, which entailed fitting a linear
model with memory score as outcome, diabetes status as exposure, adjusting for
age at cognitive assessment and sex. Next, we implemented an MR analysis of
the effects of diabetes status on cognitive score incorporating all 39 SNPs as
candidate IV using TSLS, sisVIVE, adaptive LASSO, TSHT, MR Egger, and the
proposed GENIUS approaches.

Participants were, on average, 68.1 years old (standard deviation [SD]=10.1
years old) at baseline and 1282 of them self-reported that they had diabetes
(16.7\%). The 39 SNPs jointly included in a first-stage logistic regression
model to predict diabetes status explained 3.5\% (Nagelkerke $R^{2}$) of the
variation in diabetes in the study sample, and were strongly associated as a
set with the endogenous variable (Likelihood ratio test Chi-square statistic =
162 with 39 degrees of freedom, which corresponds to a significance value
%TCIMACRO{\TEXTsymbol{<}}%
%BeginExpansion
$<$%
%EndExpansion
0.001$)$. This provides fairly compelling evidence that the IVs are not only
jointly relevant but also satisfy the first stage heteroscedasticity condition
required by MR GENIUS.

Table \ref{tab:app} shows results from both observational and IV analyses. In the
observational analysis, being diabetic was associated with an average decrease
of 0.04 points (s.e.=0.02) in memory score. MR GENIUS suggests a notably
larger diabetes-associated decrease in average memory score equal to 0.18
points (s.e.=0.14). The efficient MR GENIUS produced a similar decrease of
0.16 points (s.e.=0.14). MR-Egger produced an estimate suggesting a protective
effect of diabetes (beta=0.25, s.e.=0.35) and so did TSLS (beta=0.48,
s.e.=$\allowbreak$0.22), sisVIVE (beta=0.48) and adaptive lasso (beta=0.48,
s.e.=0.22) which gave the same point estimate, while TSHT (beta=0.45,
s.e.=0.28) gave a slightly smaller but still protective estimate. TSLS,
sisVIVE and adaptive lasso inferences coincide exactly in this application
because all 39 candidate SNPs ended up being selected as "valid" by sisVIVE
and adaptive lasso. In contrast, TSHT selected six candidate IVs only as both
valid and relevant which were therefore used to estimate the causal effect. In
conclusion, both the observational analysis and MR GENIUS found some evidence
of a harmful effect of diabetes on memory score, which supports the prevailing
hypothesis in the diabetes literature. In contrast, all other (robust and
non-robust) MR methods suggest a protective effect of diabetes on memory, a
hypothesis with little if any scientific basis in the diabetes literature.

\begin{table}[!htbp]
\caption{Estimation of $\beta_{\mbox{t2d-ms}}$, the association between type 2
diabetes and memory score.}%
\label{tab:app}
\begin{center}

\resizebox{0.6\textheight}{!}{
\begin{tabular}
[c]{rcccc}%
\hline & \multicolumn{1}{c}{$\hat{\beta}_{\mbox{t2d-ms}}$} &
\multicolumn{1}{c}{SE} & \multicolumn{1}{c}{95\% CI} &
\multicolumn{1}{p{4cm}}{\centering \# of instruments \newline selected as
invalid}\\
\hline
\multicolumn{4}{l}{Observational analysis} & \\[3pt]%
\quad\quad\quad\quad & $-0.04$ & $0.02$ & $(-0.08, 0.001)$ & -\\
\multicolumn{4}{l}{IV analyses} & \\[3pt]%
MR GENIUS & $-0.18$ & $0.14$ & $(-0.45,\phantom{0}0.08)$ & -\\
Efficient MR GENIUS & $-0.16$ & $0.14$ & $(-0.43,\phantom{0}0.11)$ & -\\
MR-Egger & $\phantom{-}0.25$ & $0.35$ & $(-0.43,\phantom{0}0.93)$ & -\\
sisVIVE & $\phantom{-}0.48$ & - & - & 0\\
TSLS & $\phantom{-}0.48$ & 0.22 & $(\phantom{-}0.05,\phantom{0}0.90)$ & -\\
Adaptive Lasso & $\phantom{-}0.48$ & - & - & 0\\
Post-adaptive Lasso & $\phantom{-}0.48$ & 0.22 &
$(\phantom{-}0.05,\phantom{0}0.90)$ & 0\\
TSHT & $\phantom{-}0.45$ & 0.28 & $(-0.10,\phantom{0}1.00)$ &
\begin{tabular}
[c]{@{}c@{}}%
0 (out of 6\\
selected as relevant)
\end{tabular}
\\[12pt]%
 &  &  &  &\\
\hline
\end{tabular}}
\end{center}
\end{table}

\section{Discussion}
\label{s:dis}

As MR gains popularity as a promising strategy to address confounding bias in
observational studies, there clearly also is a growing need for robust
MR\ methodology that relax the standard IV\ assumptions. Although a variety of
methods have recently been proposed, we have argued that MR GENIUS stands out
as an effective approach with clear advantages over other existing methods.
Whereas existing methods are technically only
consistent either as the number of candidate IVs goes to infinity (MR-Egger),
or as a majority (adaptive lasso)\ or a plurality (TSTH)\ of IVs are
valid,\ MR GENIUS\ is guaranteed to be consistent without even one valid
IV. An R package which
implements MR GENIUS is available at \url{https://github.com/bluosun/MR-GENIUS}.

In closing, we acknowledge certain limitations of MR GENIUS. First, the
approach may be vulnerable to weak IV bias which may occur if $var(A|G)$ is
weakly dependent on $G$, a possibility that was largely ruled out in this
paper. MR GENIUS is also currently not designed to handle high dimensional IVs
(where the number of IVs may exceed sample size). We plan to further develop
MR GENIUS\ to address all of these remaining challenges in future work. Doubly robust and locally efficient MR GENIUS estimation is the subject in a companion manuscript currently in preparation.

\section*{Acknowledgements}

Eric Tchetgen Tchetgen's work is funded by NIH grant R01AI104459. BaoLuo Sun's work is supported by the National University of Singapore Start-Up Grant (R-155-000-203-133). The Health
and Retirement Study genetic data are sponsored by the National Institute on
Aging (grant numbers U01AG009740, RC2AG036495, and RC4AG039029) and was
conducted by the University of Michigan. The authors thank Frank Windmeijer
for valuable discussions. \vspace*{-8pt}

\appendix

\section{Proofs of Lemmas}\label{app}

\subsection{Proof of Lemma 3.1}
\label{a1}
\begin{proof}
Under assumption 4a and taking iterated expectation with respect to $(A,G,U)$ followed by $(G,U)$,
\begin{eqnarray*}
& & E\left[  \left\{  G-E(G)\right\}  \left\{  A-E(A|G)\right\}  Y\right] \\
&  =&E\left[  \left\{  G-E(G)\right\}  \left\{  A-E(A|G)\right\}  E\left(
Y|A,G,U\right)  \right] \\
&  =&E\left[  \left\{  G-E(G)\right\}  \left\{  A-E(A|G)\right\}  \left\{
\beta_{a}(U)A+\beta_{g}(U,G)+\beta_{u}\left(  U\right)  \right\}  \right] \\
&  =&E\left[  \left\{  G-E(G)\right\}  \left\{  A-E(A|G)\right\}  \beta_{a}(U)A\right]\\
&\phantom{=}&  +E\left[  \left\{  G-E(G)\right\}  \left\{  A-E(A|G)\right\}\beta_{g}(U,G)\right] \\
&\phantom{=}&  +E\left[  \left\{  G-E(G)\right\}  \left\{  A-E(A|G)\right\}\beta_{u}\left(  U\right) \right] \\
&  =&E\left[  \left\{  G-E(G)\right\}  \left\{  1-E(A|G)\right\}  \beta_{a}(U)\{ \alpha_g(U,G)+\alpha_u(U)\}\right]\\
&\phantom{=}&  +E\left[  \left\{  G-E(G)\right\}  \left\{  \alpha_g(U,G)+\alpha_u(U)-E(A|G)\right\}\beta_{g}(U,G)\right] \\
&\phantom{=}&  +E\left[  \left\{  G-E(G)\right\}  \left\{ \alpha_g(U,G)+\alpha_u(U)-E(A|G)\right\}\beta_{u}\left(  U\right) \right] \\
&  =&E\left[  \left\{  G-E(G)\right\}  \left\{  A-E(A|G)\right\}
A E\{\beta_a(U)|G\}\right]  \\
&\phantom{=}& +E\left[  \left\{G-E(G)\right\}\left\{  1-E(A|G)\right\}\mbox{cov}\{\beta_a(U),\alpha_u(U)|G\} \right]\\
&\phantom{=}&  +E\left[  \left\{  G-E(G)\right\}  \left\{  1-E(A|G)\right\}\mbox{cov}\{\beta_a(U), \alpha_g(U,G)|G\} \right]\\
&\phantom{=}&  +E\left[  \left\{  G-E(G)\right\}\mbox{cov}\{\alpha_g(U,G),\beta_{g}\left(  U,G\right)|G\}\right]\\
&\phantom{=}& + E\left[  \left\{  G-E(G)\right\}\mbox{cov}\{\alpha_u(U),\beta_{g}\left(  U,G\right)|G\}\right]\\
&\phantom{=}&  +E\left[  \left\{  G-E(G)\right\}\mbox{cov}\{\alpha_g(U,G),\beta_{u}\left(  U\right)|G\}\right]\\
&\phantom{=}& + E\left[  \left\{  G-E(G)\right\}\mbox{cov}\{\alpha_u(U),\beta_{u}\left(  U\right)|G\}\right]
\end{eqnarray*}
Under assumption 2, $E\{\beta_a(U)|G\}=E\{\beta_a(U)\}$ and $\mbox{cov}\{\alpha_u(U),\beta_{u}\left(  U\right)|G\}=\mbox{cov}\{\alpha_u(U),\beta_{u}\left(  U\right)\}$, so that $E\left[  \left\{  G-E(G)\right\}\mbox{cov}\{\alpha_u(U),\beta_{u}\left(  U\right)|G\}\right]=0.$ Therefore, under assumption 4b, 
\begin{eqnarray*}
  \frac{E\left[  \left\{  G-E(G)\right\}  \left\{  A-E(A|G)\right\}
Y\right]  }{E\left[  \left\{  G-E(G)\right\}  \left\{  A-E(A|G)\right\}
A\right]  }=E\{\beta_{a}(U)\}%
\end{eqnarray*}
provided that $E\left[  \left\{  G-E(G)\right\}  \left\{  A-E(A|G)\right\}
A\right]=E\left[  \left\{  G-E(G)\right\}\mbox{var}(A|G)\right]  \neq0$.
\end{proof}

\subsection{Proof of Lemma 3.2}
\label{a3}

\begin{proof}
Under assumption 4* and taking iterated expectation with respect to $(A,G,U)$ followed by $(G,U)$,
\begin{eqnarray*}
& & E\left[  \left\{  G-E(G)\right\}  \left\{  A-E(A|G)\right\}  Y\right] \\
&  =&E\left[  \left\{  G-E(G)\right\}  \left\{  A-E(A|G)\right\}  E\left(
Y|A,G,U\right)  \right] \\
&  =&E\left[  \left\{  G-E(G)\right\}  \left\{  A-E(A|G)\right\}  \left\{
\beta_{a}A+\beta_{g}(U,G)+\beta_{u}\left(  U\right)  \right\}  \right] \\
&  =&E\left[  \left\{  G-E(G)\right\}  \left\{  A-E(A|G)\right\}
  A\right]\beta_a  \\
&\phantom{=}&  +E\left[  \left\{  G-E(G)\right\}\mbox{cov}\{\alpha_g(U,G),\beta_{g}\left(  U,G\right)|G\}\right]\\
&\phantom{=}& + E\left[  \left\{  G-E(G)\right\}\mbox{cov}\{\alpha_u(U),\beta_{g}\left(  U,G\right)|G\}\right]\\
&\phantom{=}&  +E\left[  \left\{  G-E(G)\right\}\mbox{cov}\{\alpha_g(U,G),\beta_{u}\left(  U\right)|G\}\right]\\
&\phantom{=}& + E\left[  \left\{  G-E(G)\right\}\mbox{cov}\{\alpha_u(U),\beta_{u}\left(  U\right)|G\}\right]
\end{eqnarray*}

The proof for lemma 3.2 follows from the observation that instead of requiring assumption 2, we just need 
$\mbox{cov}\{\alpha_u(U),\beta_{u}\left(  U\right)|G\}=\rho$
so that $$E\left[  \left\{  G-E(G)\right\}\mbox{cov}\{\alpha_u(U),\beta_{u}\left(  U\right)|G\}\right]=0,$$ and hence
\begin{eqnarray*}
  \frac{E\left[  \left\{  G-E(G)\right\}  \left\{  A-E(A|G)\right\}
Y\right]  }{E\left[  \left\{  G-E(G)\right\}  \left\{  A-E(A|G)\right\}
A\right]  }=\beta_{a}.%
\end{eqnarray*}
\end{proof}

\subsection{Proof of Lemma 4.2}
\label{a5}

\begin{proof}
The proof follows upon noting that under our assumptions,
\begin{align*}
&  \exp\left(  \varpi_{g}\left(  G\right)  \right)  E(A\exp\left(  -\varpi
_{g}\left(  G\right)  \right)  )\\
&  =E\left(  A|G\right) \\
&  =\exp\left(  \alpha_{g}\left(  G\right)  \right)  E\left(  U_{a}\right)  ,
\end{align*}
and%
\begin{align*}
&  E\left(  A|G,U\right)  -\exp\left(  \varpi_{g}\left(  G\right)  \right)
E(A\exp\left(  -\varpi_{g}\left(  G\right)  \right)  )\\
&  =\left[  U_{a}-E\left(  U_{a}\right)  \right]  \exp\left(  \alpha
_{g}\left(  G\right)  \right)  .
\end{align*}
Therefore
\begin{align*}
&  E\left[  \left\{  G-E(G)\right\}  \left\{  A\exp(-\varpi_{g}\left(
G\right)  )-E(A\exp\left(  -\varpi_{g}\left(  G\right)  \right)  )\right\}
Y\right] \\
&  =E\left[  \left\{  G-E(G)\right\}  \left\{  A\exp(-\varpi_{g}\left(
G\right)  )-E(A\exp\left(  -\varpi_{g}\left(  G\right)  \right)  )\right\}
\beta_{a}A\right] \\
&  +E\left[  \left\{  G-E(G)\right\}  \left\{  A\exp(-\varpi_{g}\left(
G\right)  )-E(A\exp\left(  -\varpi_{g}\left(  G\right)  \right)  )\right\}
\beta_{u}\left(  U\right)  \right] \\
&  +E\left[  \left\{  G-E(G)\right\}  \left\{  A\exp(-\varpi_{g}\left(
G\right)  )-E(A\exp\left(  -\varpi_{g}\left(  G\right)  \right)  )\right\}
\beta_{g}(G)\right] \\
&  =\beta_{a}E\left[  \left\{  G-E(G)\right\}  \left\{  A\exp(-\varpi
_{g}\left(  G\right)  )-E(A\exp\left(  -\varpi_{g}\left(  G\right)  \right)
)\right\}  A\right] \\
&  +\underset{=0}{\underbrace{E\left[  \left\{  G-E(G)\right\}  \left[
U_{a}-E\left(  U_{a}\right)  \right]  \beta_{u}\left(  U\right)  \right]  }}\\
&  +\underset{=0}{\underbrace{E\left[  \left\{  G-E(G)\right\}  \left\{
U_{a}-E\left(  U_{a}\right)  \right\}  \beta_{g}(G)\right]  }}\\
&  =\beta_{a}E\left[  \left\{  G-E(G)\right\}  \left\{  A-E\left(  A|G\right)
\right\}  A\exp(-\varpi_{g}\left(  G\right)  )\right] \\
&  =\beta_{a}E\left[  \left\{  G-E(G)\right\}  var(A|G)\exp(-\varpi_{g}\left(
G\right)  )\right]  ,
\end{align*}
where we used the fact that under assumption 2, $\varpi_{g}\left(  g\right)
=\alpha_{g}\left(  g\right)  ,$ therefore proving identification provided that
$var(A|G)\exp(-\varpi_{g}\left(  G\right)  )$ is a function of $G,$which holds
as long as $var\left(  A|g\right)  /var\left(  A|g=0\right)  \neq\exp\left(
\varpi_{g}\left(  g\right)  \right)  .$
\end{proof}

\subsection{Proof of Lemma 4.3}
\label{a6}

\begin{proof}
We first note that for any additive function $t(A,G)=t_{1}(A)+t_{2}(G),$
\[
E\left(  t(A,G)\left\{  G-E(G|A=0)\right\}  \left\{  A-E(A|G=0)\right\}
\exp\left\{  -\varphi_{g}\left(  G\right)  A\right\}  \right)  =0
\]
because
\begin{align*}
&  E\left(  t(A,G)\left\{  G-E(G|A=0)\right\}  \left\{  A-E(A|G=0)\right\}
\exp\left\{  -\varphi_{g}\left(  G\right)  A\right\}  \right) \\
&  =\sum_{a,g}f(a,g)t(a,g)\left\{  g-E(G|A=0)\right\}  \left\{
a-E(A|G=0)\right\}  \exp\left\{  -\varphi_{g}\left(  g\right)  a\right\} \\
&  \propto\sum_{a,g} \left\{  f(a|g=0)f(g|a=0)\exp\left\{  \varphi_{g}\left(
g\right)  a\right\}  t(a,g) \right. \\
& \phantom{\propto\sum} \left.  \times\left\{  g-E(G|A=0)\right\}  \left\{
a-E(A|G=0)\right\} \exp\left\{  -\varphi_{g}\left(  g\right)  a\right\}
\right\} \\
&  =\sum_{a,g}f(a|g=0)f(g|a=0)t(a,g)\left\{  g-E(G|A=0)\right\}  \left\{
a-E(A|G=0)\right\} \\
&  =0
\end{align*}
where we used the fact that
\[
f(a,g)\propto(a|g=0)f(g|a=0)\exp\left\{  \varphi_{g}\left(  g\right)
a\right\}  ,
\]
see for example \cite{tchetgen2009doubly}. It is straightforward to
verify that the
\[
\theta=-\ln\left(  1-\frac{E\left[  \left\{  G-E(G|A=0)\right\}  \left\{
A-E(A|G=0)\right\}  Y\exp\left\{  -\varphi_{g}\left(  G\right)  A\right\}
\right]  }{E\left[  \left\{  G-E(G|A=0)\right\}  \left\{  A-E(A|G=0)\right\}
AY\exp\left\{  -\varphi_{g}\left(  G\right)  A\right\}  \right]  }\right)
\]
Next,
\begin{align*}
&  E\left[  \left\{  G-E(G|A=0)\right\}  \left\{  A-E(A|G=0)\right\}
Y\exp\left\{  -\varphi_{g}\left(  G\right)  A\right\}  \right]  \\
&  =E\left[  \left\{  G-E(G|A=0)\right\}  \left\{  A-E(A|G=0)\right\}
\exp\left(  \beta_{a}A\right)  E\left(  Y|A=0,G,U\right)  \exp\left\{
-\varphi_{g}\left(  G\right)  A\right\}  \right] \\
&  =E\left[  \left\{  G-E(G|A=0)\right\}  \left\{  A-E(A|G=0)\right\}  \left(
\exp\left(  \beta_{a}\right)  -1\right)  AE\left(  Y|A=0,G,U\right)
\exp\left\{  -\varphi_{g}\left(  G\right)  A\right\}  \right] \\
&  +E\left[  \left\{  G-E(G|A=0)\right\}  \left\{  A-E(A|G=0)\right\}
E\left(  Y|A=0,G,U\right)  \exp\left\{  -\varphi_{g}\left(  G\right)
A\right\}  \right] \\
&  =\left(  \exp\left(  \beta_{a}\right)  -1\right)  E\left[  \left\{
G-E(G|A=0)\right\}  \left\{  A-E(A|G=0)\right\}  AE\left(  Y|A=0,G,U\right)
\exp\left\{  -\varphi_{g}\left(  G\right)  A\right\}  \right] \\
&  +E\left[  \left\{  G-E(G|A=0)\right\}  \left\{  A-E(A|G=0)\right\}  \left(
E\left(  Y|A=0,G,U\right)  -E\left(  Y|A=0,G=0,U\right)  \right) \right. \\
& \left.  \phantom{=}\times\exp\left\{ -\varphi_{g}\left(  G\right)
A\right\}  \right] \\
&  +E\left[  \left\{  G-E(G|A=0)\right\}  \left\{  A-E(A|G=0)\right\}  \left(
E\left(  Y|A=0,G=0,U\right)  \right)  \exp\left\{  -\varphi_{g}\left(
G\right)  A\right\}  \right] \\
&  =\left(  \exp\left(  \beta_{a}\right)  -1\right)  E\left[  \left\{
G-E(G|A=0)\right\}  \left\{  A-E(A|G=0)\right\}  AE\left(  Y|A=0,G,U\right)
\exp\left\{  -\varphi_{g}\left(  G\right)  A\right\}  \right] \\
&  +\underset{=0}{\underbrace{E\left[  \left\{  G-E(G|A=0)\right\}  \left\{
A-E(A|G=0)\right\}  \beta_{g}\left(  G\right)  \exp\left\{  -\varphi
_{g}\left(  G\right)  A\right\}  \right]  }}\\
&  +\underset{=0}{\underbrace{E\left[  \left\{  G-E(G|A=0)\right\}  \left\{
A-E(A|G=0)\right\}  E\left[  E\left(  Y|A=0,G=0,U\right)  |A\right]
\exp\left\{  -\varphi_{g}\left(  G\right)  A\right\}  \right]  }}%
\end{align*}
Likewise%
\begin{align*}
&  E\left[  \left\{  G-E(G|A=0)\right\}  \left\{  A-E(A|G=0)\right\}
AY\exp\left\{  -\varphi_{g}\left(  G\right)  A\right\}  \right] \\
&  =E\left[  \left\{  G-E(G|A=0)\right\}  \left\{  A-E(A|G=0)\right\}
\exp\left(  \beta_{a}\right)  E(Y|A=0,G,U)A\exp\left\{  -\varphi_{g}\left(
G\right)  A\right\}  \right]
\end{align*}
Therefore
\begin{align*}
&  \frac{E\left[  \left\{  G-E(G|A=0)\right\}  \left\{  A-E(A|G=0)\right\}
Y\exp\left\{  -\varphi_{g}\left(  G\right)  A\right\}  \right]  }{E\left[
\left\{  G-E(G|A=0)\right\}  \left\{  A-E(A|G=0)\right\}  AY\exp\left\{
-\varphi_{g}\left(  G\right)  A\right\}  \right]  }\\
&  =\frac{\left(  \exp\left(  \beta_{a}\right)  -1\right)  E\left[  \left\{
G-E(G|A=0)\right\}  \left\{  A-E(A|G=0)\right\}  AE\left(  Y|A=0,G,U\right)
\exp\left\{  -\varphi_{g}\left(  G\right)  A\right\}  \right]  }{\exp\left(
\beta_{a}\right)  E\left[  \left\{  G-E(G|A=0)\right\}  \left\{
A-E(A|G=0)\right\}  E(Y|A=0,G,U)A\exp\left\{  -\varphi_{g}\left(  G\right)
A\right\}  \right]  }\\
&  =\frac{\left(  \exp\left(  \beta_{a}\right)  -1\right)  \ }{\exp\left(
\beta_{a}\right)  \ }%
\end{align*}%
\begin{align*}
\theta &  =-\ln\left(  1-\frac{\left(  \exp\left(  \beta_{a}\right)
-1\right)  \ }{\exp\left(  \beta_{a}\right)  \ }\right) \\
&  =-\ln\exp\left(  -\beta_{a}\right)  \ \\
&  =\beta_{a}%
\end{align*}
provided that%
\begin{align*}
&  E\left[  \left\{  G-E(G|A=0)\right\}  \left\{  A-E(A|G=0)\right\}
E(Y|A=0,G,U)A\exp\left\{  -\varphi_{g}\left(  G\right)  A\right\}  \right] \\
&  =E\left[  \left\{  G-E(G|A=0)\right\}  \left\{  A-E(A|G=0)\right\}
\beta_{g}\left(  G\right)  A\exp\left\{  -\varphi_{g}\left(  G\right)
A\right\}  \right] \\
&  \neq0
\end{align*}
which holds by assumption because $\gamma_{ag}(g)$ $=\left(  \exp\left(
\beta_{a}\right)  -1\right)  \beta_{g}\left(  g\right)  .$
\end{proof}

\subsection{Proof of Lemma 4.4}

\begin{proof}
We note that by assumption
\[
E\left(  dN(y)-d\mathbf{B}_{a}\left(  y\right)  A-d\mathbf{B}_{g}\left(
y\right)  G|R(y)=1,A,G,U\right)  =d\mathbf{B}_{0}\left(  y\right)
+d\mathbf{B}_{u}\left(  y,U\right)  ,
\]
and
\begin{align*}
&  E(\exp\left\{  \mathbf{B}_{a}\left(  y\right)  A+\mathbf{B}_{g}\left(
y\right)  G\right\}  R(y)|A,G,U)\\
&  =\exp\left\{  -\mathbf{B}_{0}\left(  y\right)  -\mathbf{B}_{u}\left(
y,U\right)  \right\}  .
\end{align*}
Therefore
\begin{align*}
&  E\left\{  W\left(  y,\mathbf{B}_{a}\left(  y\right)  ,\mathbf{B}_{g}\left(
y\right)  \right)  \right\} \\
&  =E\left\{  \left(  d\mathbf{B}_{0}\left(  y\right)  +d\mathbf{B}_{u}\left(
y,U\right)  \right)  \exp\left\{  -\mathbf{B}_{0}\left(  y\right)
-\mathbf{B}_{u}\left(  y,U\right)  \right\}  \left(
\begin{array}
[c]{c}%
\left(  G-E(G)\right) \\
\left(  G-E(G)\right)  \left(  A-E(A|G)\right)
\end{array}
\right)  \right\} \\
&  =E\left\{  \left(
\begin{array}
[c]{c}%
0\\
\left(  d\mathbf{B}_{0}\left(  y\right)  +d\mathbf{B}_{u}\left(  y,U\right)
\right)  \exp\left\{  -\mathbf{B}_{0}\left(  y\right)  -\mathbf{B}_{u}\left(
y,U\right)  \right\}  \left(  U-E(U)\right)  \left(  G-E(G)\right)
\end{array}
\right)  \right\} \\
&  =0.
\end{align*}

\end{proof}

\section{Variance Estimation}

\subsection{Single IV}
\label{a2}
The estimating equation in (3.2) involves the estimated nuisance parameters
$\hat{\mu}=\mathbb{P}_{n}(G)$ and $\hat{\psi}$ of the model $E(A|G;\psi)$. To
account for the effect of nuisance parameter estimation on the subsequent
estimation of $\beta_{a}$, the empirical moment conditions are stacked to
form
\[
m_{\theta}(\theta)=\mathbb{P}_{n}%
\begin{bmatrix}
G-\mu\\
(1,G)^{\prime}\left[  A-E(A|G;\psi)\right] \\
(G-\mu)\left[  A-E(A|G;\psi)\right]  (Y-\beta_{a}A)
\end{bmatrix}
,\text{ where }\theta=(\mu,\psi,\beta_{a}).
\]
The estimation procedure satisfies the joint conditions $m_{\theta}\left(
\hat{\theta}\right)  =0$. Without loss of generality, we specify $\left[
A-E(A|G;\psi_{0})\right]  $ as a main effects model with intercept. Assume
standard regularity conditions and expand $\hat{\theta}$ around the true
parameter value $\theta_{0}$ yields
\[
\sqrt{n}\left(  \hat{\theta}-\theta_{0}\right)  =-\left[  \frac{\partial
{m}_{\theta}\left(  {\theta}\right)  }{\partial\theta}\bigg\rvert_{\theta
^{\ast}}\right]  ^{-}\sqrt{n}{m}_{\theta}\left(  {\theta}_{0}\right)  ,
\]
where $\theta^{\ast}$ is intermediate in value between $\hat{\theta}$ and
$\theta_{0}$. It follows that
\begin{align*}
\sqrt{n}{m}_{\theta}\left(  {\theta}_{0}\right)   &  =\sqrt{n}\mathbb{P}_{n}%
\begin{bmatrix}
G-\mu_{0}\\
(1,G)^{\prime}\left[  A-E(A|G;\psi_{0})\right] \\
(G-\mu_{0})\left[  A-E(A|G;\psi_{0})\right]  (Y-\beta_{a0}A)
\end{bmatrix}
\\
&  =\sqrt{n}\mathbb{P}_{n}\left\{  \tilde{m}(\theta_{0})\right\}
\overset{d}{\rightarrow}N(0,E\left[  \tilde{m}(\theta_{0})\tilde{m}(\theta
_{0})^{\prime}\right]  ),
\end{align*}
while for the "bread" matrix
\begin{align*}
&  \frac{\partial{m}_{\theta}\left(  {\theta}\right)  }{\partial\theta
}\bigg\rvert_{\theta^{\ast}}=B^{\ast}(\theta^{\ast})=\\
&  \mathbb{P}_{n}%
\begin{bmatrix}
-1 & 0_{1\times2} & 0\\
0_{2\times1} & -\left\{  (1,G)^{\prime}\frac{\partial}{\partial\psi}%
E(A|G;\psi)\bigg\rvert_{\psi^{\ast}}\right\}  & 0_{2\times1}\\
& \left\{  \frac{\partial\widehat{U}}{\partial\mu}\bigg\rvert_{\mu^{\ast}%
},\frac{\partial\widehat{U}}{\partial\psi}\bigg\rvert_{\psi^{\ast}}%
,\frac{\partial\widehat{U}}{\partial\beta_{a}}\bigg\rvert_{\beta_{a}^{\ast}%
}\right\}  &
\end{bmatrix}
,
\end{align*}
where
\[
\frac{\partial}{\partial\psi}E(A|G;\psi)=%
\begin{cases}
(1,G), & \text{for continuous }A\ \\
\frac{\exp{(1,G)\psi}}{1+\exp{(1,G)\psi}}\left(  1-\frac{\exp{(1,G)\psi}%
}{1+\exp{(1,G^{\prime})\psi}}\right)  (1,G), & \text{for binary A (logit
model),}%
\end{cases}
\]
and
\begin{align*}
\frac{\partial\widehat{U}}{\partial\mu}  &  =-(A-E(A|G;\psi))(Y-\beta_{a}A)\\
\frac{\partial\widehat{U}}{\partial\psi}  &  =-(G-\mu)(Y-\beta_{a}%
A)\frac{\partial}{\partial\psi}E(A|G;\psi)\\
\frac{\partial\widehat{U}}{\partial\beta_{a}}  &  =-(G-\mu)(A-E(A|G;\psi))A.
\end{align*}
Assume that the matrix $B(\theta_{0})$ is non-singular, where the entries in
$B(\theta_{0})$ are the expected values of the sample averages in $B^{\ast
}(\theta^{\ast})$, evaluated at $\theta_{0}$. Then $B^{\ast}(\theta^{\ast
})\overset{p}{\rightarrow}B(\theta_{0})$, and
\begin{align}
\label{s1}
&  \sqrt{n}\left(  \hat{\theta}-\theta_{0}\right)  \overset{d}{\rightarrow
}\nonumber\\
&  N\left(  0,B(\theta_{0})^{-}E\left[  \tilde{m}(\theta_{0})\tilde{m}%
(\theta_{0})^{\prime}\right]  B(\theta_{0})^{-\prime}\right)  . \tag{S1}%
\end{align}
Replacing the expected values in (\ref{s1}) with sample averages evaluated at
$\hat{\theta}$ yields a consistent estimator of the asymptotic covariance
matrix. For inference about $\beta_{a}$, one may report its Wald-type $95\%$
confidence interval constructed with the corresponding component of the
estimated covariance matrix for $\hat{\theta}$.

\subsection{Multiple IVs}
\label{a4}

Let $\widehat{\beta}_{a}$ be the solution to (4.2) with optimal weight
$\widehat{W}_{opt}=$ $\mathbb{P}_{n}\left[  \widehat{U}\left(  {\beta}%
_{a}\right)  \widehat{U}\left(  {\beta}_{a}\right)  ^{\prime}\right]  ^{-}$
where $T^{-}$ denotes the generalized inverse of matrix $T$. The empirical
moment conditions $\widehat{U}\left(  \beta_{a}\right)  $ in (4.2) involves the
first stage estimates $\hat{\mu}=\mathbb{P}_{n}G$ as well as $\hat{\psi}$ of
the model $E(A|G; \psi)$, which effects need to be accounted for in the
subsequent estimation of $\beta_{a}$. Without loss of generality, we specify
$\left[  A-E(A|G;\psi_{0})\right]  $ as a main effects model with intercept.
If there are $k$ IVs, let
\begin{align*}
m_{\mu}(\mu)  &  =\mathbb{P}_{n}(G-\mu)\\
m_{\psi}(\psi)  &  =\mathbb{P}_{n}(1,G^{\prime})^{\prime}[A-E(A|G;\psi)]
\end{align*}
be the $k$ and $(k+1)$ empirical moment conditions of obtaining $\left(
\hat{\mu},\hat{\psi}\right)  $ respectively. For iterated or continuously
updated GMM procedures in which $\beta_{a}$ is estimated simultaneously with
the optimal weight, the first order condition of (\ref{GMM}) is
\begin{align*}
m_{\beta_{a}}(\beta_{a})=\left\{  \mathbb{P}_{n}\left[  \frac{\partial
\widehat{U}\left(  \beta_{a}\right)  }{\partial\beta_{a}}\right]  \right\}
^{\prime}\widehat{W}_{opt}(\beta_{a}) \mathbb{P}_{n}\left[  \widehat{U}\left(
\beta_{a}\right)  \right]  +o_{p}\left(  n^{-1/2}\right)  .
\end{align*}
The two-stage procedure solution satisfies the joint moment conditions
\[
{m}_{\theta}\left(  \hat{\theta}\right)  =\left(  m_{\mu}\left(  \hat{\mu
}\right)  ,m_{\psi}\left(  \hat{\psi}\right)  ,m_{\beta_{a}}\left(  \hat
{\beta}_{a}\right)  \right)  ^{\prime}=0, \phantom{00}\hat{\theta}=\left(
\hat{\mu},\hat{\psi},\hat{\beta}_{a}\right)  .
\]
Assume standard regularity conditions and expand $\hat{\theta}$ around the
true parameter value $\theta_{0}$ yields
\[
\sqrt{n}\left(  \hat{\theta}-\theta_{0}\right)  =- \left[  \frac{\partial
{m}_{\theta}\left(  {\theta}\right)  }{\partial\theta}\bigg\rvert_{\theta
^{\ast}} \right]  ^{-}\sqrt{n}{m}_{\theta}\left(  {\theta}_{0}\right)  ,
\]
where $\theta^{\ast}$ is intermediate in value between $\hat{\theta}$ and
$\theta_{0}$. Consider
\begin{align*}
&  \sqrt{n}{m}_{\theta}\left(  {\theta}_{0}\right)  =\\
&
\begin{bmatrix}
I_{(2k+1)\times(2k+1)} & 0_{(2k+1)\times k}\\
0_{1\times(2k+1)} & \left\{  \mathbb{P}_{n}\left[  \frac{\partial
\widehat{U}\left(  \beta_{a}\right)  }{\partial\beta_{a}}\bigg\rvert_{\beta
_{a0}}\right]  \right\}  ^{\prime}\widehat{W}_{opt}(\beta_{a0})
\end{bmatrix}
\sqrt{n}\mathbb{P}_{n}
\begin{bmatrix}
G-\mu_{0}\\
(1,G^{\prime})^{\prime}\left[  A-E(A|G;\psi_{0})\right] \\
U(\beta_{a0})
\end{bmatrix}
+o_{p}(1).
\end{align*}

Let
\[
\Lambda=E\left(  \frac{\partial{U}\left(  \beta_{a}\right)  }{\partial
\beta_{a}}\bigg\rvert_{\beta_{a0}}\right)  ,\phantom{00}\Omega=E\left[
{U}\left(  {\beta}_{a0}\right)  U\left(  {\beta}_{a0}\right)  ^{\prime
}\right]  ,
\]
so that
\[
\left\{  \mathbb{P}_{n}\left[  \frac{\partial\widehat{U}\left(  \beta
_{a}\right)  }{\partial\beta_{a}}\bigg\rvert_{\beta_{a0}}\right]  \right\}
^{\prime}\overset{p}{\rightarrow}\Lambda^{\prime},\phantom{00}\widehat{W}%
_{opt}(\beta_{a0})\overset{p}{\rightarrow}\Omega^{-}.
\]
Then
\begin{align*}
\sqrt{n}  &  \mathbb{P}_{n}%
\begin{bmatrix}
G-\mu_{0}\\
(1,G^{\prime})^{\prime}\left[  A-E(A|G;\psi_{0})\right] \\
U(\beta_{a0})
\end{bmatrix}
=\sqrt{n}\mathbb{P}_{n}\left\{  \tilde{m}(\theta_{0})\right\} \\
&  \overset{d}{\rightarrow}N(0,E\left[  \tilde{m}(\theta_{0})\tilde{m}%
(\theta_{0})^{\prime}\right]  ),
\end{align*}
and by Slutsky's theorem
\begin{align*}
\sqrt{n}{m}_{\theta}\left(  {\theta}_{0}\right)   &  \overset{d}{\rightarrow}%
\begin{bmatrix}
I_{(2k+1)\times(2k+1)} & 0_{(2k+1)\times k}\\
0_{1\times(2k+1)} & \Lambda^{\prime}\Omega^{-}%
\end{bmatrix}
N(0,E\left[  \tilde{m}(\theta_{0})\tilde{m}(\theta_{0})^{\prime}\right]  )\\
&  =M(\theta_{0})N(0,E\left[  \tilde{m}(\theta_{0})\tilde{m}(\theta
_{0})^{\prime}\right]  ).
\end{align*}
Next consider the "bread" matrix
\begin{align*}
&  \frac{\partial{m}_{\theta}\left(  {\theta}\right)  }{\partial\theta
}\bigg\rvert_{\theta^{\ast}}=B^{\ast}(\theta^{\ast})=\\
&
\begin{bmatrix}
-I_{k\times k} & 0_{k\times(k+1)} & 0_{k\times1}\\
0_{(k+1)\times k} & -\mathbb{P}_{n}\left\{  (1,G^{\prime})^{\prime}%
\frac{\partial}{\partial\psi}E(A|G;\psi)\bigg\rvert_{\psi^{\ast}}\right\}  &
0_{(k+1)\times1}\\
& \left\{  \mathbb{P}_{n}\left[  \frac{\partial\widehat{U}\left(  \beta
_{a}\right)  }{\partial\beta_{a}}\bigg\rvert_{\beta_{a}^{\ast}}\right]
\right\}  ^{\prime}\widehat{W}_{opt}(\beta_{a}^{\ast})\mathbb{P}_{n}\left\{
\frac{\partial\widehat{U}}{\partial\mu}\bigg\rvert_{\mu^{\ast}},\frac
{\partial\widehat{U}}{\partial\psi}\bigg\rvert_{\psi^{\ast}},\frac
{\partial\widehat{U}}{\partial\beta_{a}}\bigg\rvert_{\beta_{a}^{\ast}%
}\right\}  +o_{p}(1) &
\end{bmatrix}
,
\end{align*}
where
\[
\frac{\partial}{\partial\psi}E(A|G;\psi)=%
\begin{cases}
(1,G^{\prime}), & \text{for continuous }A\ \\
\frac{\exp{(1,G^{\prime})\psi}}{1+\exp{(1,G^{\prime})\psi}}\left(
1-\frac{\exp{(1,G^{\prime})\psi}}{1+\exp{(1,G^{\prime})\psi}}\right)
(1,G^{\prime}), & \text{for binary A (logit model),}%
\end{cases}
\]
and
\begin{align*}
\frac{\partial\widehat{U}}{\partial\mu}  &  =-I_{k\times k}(A-E(A|G;\psi
))(Y-\beta_{a}A)\\
\frac{\partial\widehat{U}}{\partial\psi}  &  =-(G-\mu)(Y-\beta_{a}%
A)\frac{\partial}{\partial\psi}E(A|G;\psi)\\
\frac{\partial\widehat{U}}{\partial\beta_{a}}  &  =-(G-\mu)(A-E(A|G;\psi))A.
\end{align*}
Assume that the matrix $B(\theta_{0})$ is non-singular, where the entries in
$B(\theta_{0})$ are the expected values of the sample averages in $B^{\ast
}(\theta^{\ast})$, evaluated at $\theta_{0}$. Then $B^{\ast}(\theta^{\ast
})\overset{p}{\rightarrow}B(\theta_{0})$, and
\begin{align}
\label{s4}
&  \sqrt{n}\left(  \hat{\theta}-\theta_{0}\right)  \overset{d}{\rightarrow
}\nonumber\\
&  N\left(  0,B(\theta_{0})^{-}M(\theta_{0})E\left[  \tilde{m}(\theta
_{0})\tilde{m}(\theta_{0})^{\prime}\right]  M(\theta_{0})^{\prime}B(\theta
_{0})^{-\prime}\right) . \tag{S4}%
\end{align}
In practice, replacing the expected values in (\ref{s4}) with sample averages
evaluated at $\hat{\theta}$ yields a consistent estimator of the asymptotic
covariance matrix. In addition, centering the IV moment conditions
$\widehat{U}(\beta_{a})$ when estimating the covariance matrix $E\left[
\tilde{m}(\theta_{0})\tilde{m}(\theta_{0})^{\prime}\right]  $ may improve
finite sample inference. For inference about $\beta_{a}$, one may report its
Wald-type $95\%$ confidence interval constructed with the corresponding
component of the estimated covariance matrix for $\hat{\theta}$. The above
variance estimation framework can accommodate baseline covariates $C$ by
stacking the moment conditions for $\hat{E}(G|C)$ and $\hat{E}(A|G,C)$
instead, as described in estimating equation (4.1).

\bibliographystyle{imsart-nameyear}
\bibliography{refs}

\end{document}